\documentstyle[prb,aps,float,twocolumn,epsf]{revtex}

\begin{document}
\draft
\title{Theory of coherent acoustic phonons in InGaN/GaN multi-quantum wells}

\author{G. D. Sanders and C. J. Stanton}
\address{Department of Physics, University of Florida, Box 118440\\
Gainesville, Florida 32611-8440}

\author{Chang Sub Kim}
\address{Department of Physics, Chonnam National University, \\
Kwangju, 500-757, Korea}

\date{\today}

\maketitle

\begin{abstract}
A microscopic theory for the generation and propagation of coherent LA
phonons in  pseudomorphically strained wurtzite (0001) $InGaN/GaN$
multi-quantum well (MQW) $pin$ diodes is presented. 
The generation of coherent LA phonons is driven by photoexcitation of
electron-hole pairs by an ultrafast Gaussian pump laser and is treated
theoretically using the density matrix formalism.
We use realistic wurtzite bandstructures taking valence-band mixing
and strain-induced piezoelectric fields into account. In addition,
the many-body Coulomb interaction is treated in the screened time-dependent
Hartree-Fock approximation.
We find that under typical experimental conditions, our
microscopic theory can be simplified and mapped onto a loaded string
problem which can be easily solved.
\end{abstract}


\section{Introduction}

In recent years, experiments have shown that
optical excitation of electron-hole
pairs in semiconductors by ultrafast lasers 
can coherently excite longitudinal optical phonon modes in semiconductors.
\cite{cho90,kutt92,pfeifer92,cheng91,albricht92,kuznetsov94,kuznetsov95,thoen98,yamamoto94,bartels99}
In uniform bulk semiconductors, since the laser wavelength is much larger than
the lattice spacing, the photo-generated carriers are typically excited
by the optical pump over spatial areas that are much larger than the
lattice unit cell.  As a result, the excited carrier populations are
generated in a macroscopic state and the carrier density matrix has
only a $q \approx 0$ Fourier component.  Coupling of the photoexcited
carriers to the phonons leads only to coherent optical phonon modes
with $q \approx 0$.  Since the frequency of the $q \approx 0$ acoustic
phonon is zero, coherent acoustic phonons 
are not excited in bulk semiconductors.

In semiconductor superlattices, even though the laser pump has a
wavelength large compared to the lattice spacing, the pump can
preferentially generate electron-hole pairs in the wells.  The
result is  to create photoexcited carrier distributions that have the
periodicity of the superlattice.  Since the density matrix of the 
photo-excited carrier populations now has a $q \ne 0$ Fourier 
component, the photo-excited carriers can not only couple to
the optical phonon modes, but they can also generate coherent acoustic
phonon modes with a nonzero frequency and wavevector 
$q \approx 2 \pi /L$ where $L$ is the superlattice
period.  In superlattices, the coherent phonon oscillation of zone
folded acoustic phonons has been observed in $AlAs/GaAs$ superlattices.
\cite{yamamoto94,bartels99}. However, the reflection modulation, observed
to be on the order of $\Delta R/R \sim 10^{-5}-10^{-6}$, is very small.
\cite{bartels99}

Recently, C.-K. Sun et. al. \cite{sun00} reported studies of coherent
acoustic phonon oscillations in wurtzite (0001) $InGaN/GaN$
multi-quantum well samples with strain induced piezoelectric fields.
Owing to the strong piezoelectric fields at the interfaces, huge
coherent acoustic phonon oscillations were observed.  The oscillations
were strong enough to be seen in the transmision (rather than the usual
reflectivity) with $\Delta T/T \sim 10^{-2}-10^{-3}$.  The oscillation
frequency, in the THz range, corresponding to the LA phonon frequency
with $q \approx 2 \pi /L$, varied between samples in accordance with
their different superlattice periods, $L$.

In this paper, we formulate a microscopic model for the generation of
coherent acoustic phonons in strained wurtzite superlattices via
ultrafast laser photo-excitation of real carriers. Whereas in bulk systems
the microscopic theory of coherent LO phonons can be mapped onto a forced
oscillator model \cite{kuznetsov94}, we show that coherent LA phonon generation
in superlattices, under appropriate conditions, can be mapped onto a
\textit{loaded string model} which is readily solved for the lattice 
displacement. Since acoustic phonons are almost the same in the well as in 
the barrier, to lowest order we can treat the string as being uniform
\cite{yu96}. The forcing term on the string, however, is not uniform since
photoexcitation of carriers occurs only in the wells.

Our paper thus provides justification for using a simple, uniform string model
with a nonuniform forcing term, rather than a more complicated microscopic
theory. In addition, we provide a microscopic expression for the forcing
term to use in the simplified string model.
The string model provides additional insight into the physics of the coherent
LA phonons.

\section{Microscopic theory}

In this section, we derive the microscopic theory for coherent acoustic
phonon generation in superlattices and multiple quantum wells, including the
effects of (i) bandstructure, (ii) strain, (iii) piezoelectric fields,
(iv) Coulomb interactions, and (v) laser optical excitation.
In section III, we will show how this reduces to a simplified driven uniform
string model with a nonuniform forcing term and a microscopic expression for
the forcing term.

We model photogeneration of electrons and holes and
the subsequant excitation of coherent acoustic phonons in a multi-quantum
well (MQW) $pin$ diode shown schematically in Fig.\ \ref{diode}.
The intrinsic active region consists of 
a left $GaN$ buffer region, several pseudomorphically strained
(0001) $In_{x}Ga_{1-x}N$ quantum wells sandwiched between $GaN$ barriers,
and a right $GaN$ buffer region as indicated in the figure.
The $P$ and $N$ regions are assumed to be abruptly terminated $p$- and
$n$-doped $GaN$ bulk layers separated by a distance, $L$, across which a
voltage drop, $\Delta V = V_A$, is maintained.
Photoexcitation of carriers is achieved by means of an ultrafast laser
pulse incident normally along the (0001) growth direction, taken to
coincide with the $z$-axis.

\subsection{Bulk Bandstructure}
In bulk systems, the conduction and valence bands 
in wurzite crystals including the effects of
strain are treated using effective mass theory. Near the band edge, the
effective mass Hamiltonian for electrons is described by a $2 \times 2$
matrix which depends explicitly on electron wavevector, {\bf k}, and the strain
tensor, {\bf $\epsilon$}. The electron Bloch basis states are taken to be
\begin{mathletters}
\label{cBasis}
\begin{equation}
\arrowvert c,1 \rangle = \arrowvert S \uparrow \rangle
\end{equation}
\begin{equation}
\arrowvert c,2 \rangle = \arrowvert S \downarrow \rangle .
\end{equation}
\end{mathletters}
The conduction band Hamiltonian is diagonal and we have (relative to the
bottom of the conduction band) (Refs.
\onlinecite{jeon97} and \onlinecite{chuang96b}) 
\begin{eqnarray}
H^{c}_{2 \times 2}({\bf k}, {\bf \epsilon}) = \{ 
\frac{\hbar^{2} k_{z}^{2}}{2 m^{*}_{z}} + 
\frac{\hbar^{2} k_{t}^{2}}{2 m^{*}_{xy}} \nonumber \\
 + \ a_{c,z} \epsilon_{zz} + a_{c,xy} (\epsilon_{xx} + \epsilon_{yy}) \}
 \ {\bf I}_{2 \times 2} .
\label{HcMatrix}
\end{eqnarray}
where ${\bf I}_{2 \times 2}$ is the identity matrix.
The electron effective masses along z (taken parallel to the c-axis) and in
the $xy$ plane are $m^*_z$ and $m^*_{xy}$, respectively, $k_t^2=k_x^2+k_y^2$,
and $\epsilon_{xx}$, $\epsilon_{yy}$ and $\epsilon_{zz}$ are strain tensor
components, and $a_{c,z}$ and $a_{c,xy}$ are the deformation potentials.

The Hamiltonian for the valence bands is a $6 \times 6$ matrix.
Following Ref. \onlinecite{chuang96a}, the Hamiltonian (relative to the
top of the valence band) can
be block diagonalized into two degnerate $3 \times 3$ submatrices if
we adopt the Bloch basis states
\begin{mathletters}
\label{vBasis}
\begin{equation}
 \arrowvert v,1 \rangle =
- \frac{\alpha^{*}}{\sqrt{2}} \arrowvert (X + i Y) \uparrow \rangle
+ \frac{\alpha}{\sqrt{2}} \arrowvert (X - i Y) \downarrow \rangle
\end{equation}
\begin{equation}
\arrowvert v,2 \rangle =
  \frac{\beta}{\sqrt{2}} \arrowvert (X - i Y) \uparrow \rangle
- \frac{\beta^{*}}{\sqrt{2}} \arrowvert (X + i Y) \downarrow \rangle
\end{equation}
\begin{equation}
\arrowvert v,3 \rangle =
  \beta^{*} \arrowvert Z \uparrow \rangle
+ \beta \arrowvert Z \downarrow \rangle
\end{equation}
\begin{equation}
\arrowvert v,4 \rangle =
- \frac{\alpha^{*}}{\sqrt{2}} \arrowvert (X + i Y) \uparrow \rangle
- \frac{\alpha}{\sqrt{2}} \arrowvert (X - i Y) \downarrow \rangle
\end{equation}
\begin{equation}
\arrowvert v,5 \rangle =
  \frac{\beta}{\sqrt{2}} \arrowvert (X - i Y) \uparrow \rangle
+ \frac{\beta^{*}}{\sqrt{2}} \arrowvert (X + i Y) \downarrow \rangle
\end{equation}
\begin{equation}
\arrowvert v,6 \rangle =
- \beta^{*} \arrowvert Z \uparrow \rangle
+ \beta \arrowvert Z \downarrow \rangle
\end{equation}
\end{mathletters}
The phase factors, $\alpha$ and $\beta$, are functions of the angle
$\phi = \tan^{-1} (k_{y}/k_{x})$ and are given by
\begin{mathletters}
\label{phases}
\begin{equation}
\alpha(\phi) = \frac{1}{\sqrt{2}} \ e^{i(3 \pi /4 + 3 \phi /2) }
\end{equation}
\begin{equation}
\beta(\phi)  = \frac{1}{\sqrt{2}} \ e^{i(  \pi /4 +   \phi /2) } .
\end{equation}
\end{mathletters} 
The block diagonalized Hamiltonian can be written as
\begin{equation}
H^{v}_{6 \times 6}({\bf k}, {\bf \epsilon}) =
\left(
\begin{array}{cc}
     H^{U}_{3 \times 3}( {\bf k}, {\bf \epsilon} )  & 0 \\
     0  & H^{L}_{3 \times 3}( {\bf k}, {\bf \epsilon} )
\end{array}
\right) ,
\label{HvMatrix}
\end{equation}
where the upper and lower blocks of the Hamiltonian are
\begin{mathletters}
\label{HulMatrix}
\begin{equation}
H^{U}_{3 \times 3}( {\bf k}, {\bf \epsilon} ) =
\left(
\begin{array}{ccc}
F      & K_t         & -iH_t         \\
K_t    & G           & \Delta-iH_t   \\
iH_t   & \Delta+iH_t & \lambda
\end{array}
\right) 
\end{equation}
and
\begin{equation}
H^{L}_{3 \times 3}( {\bf k}, {\bf \epsilon} ) =
\left(
\begin{array}{ccc}
F      & K_t         &  iH_t         \\
K_t    & G           & \Delta+iH_t   \\
-iH_t  & \Delta-iH_t & \lambda
\end{array}
\right) . 
\end{equation}
\end{mathletters}

The elements appearing in the $3 \times 3$ Hamiltonian matrices are
\begin{mathletters}
\label{ValenceMatrixElements} 
\begin{equation}
F = \Delta_1 + \Delta_2 + \lambda + \theta
\end{equation}
\begin{equation}
G = \Delta_1 - \Delta_2 + \lambda + \theta
\end{equation}
\begin{equation}
K_t = \frac{\hbar^2}{2 m_0} A_5 k_t^2
\end{equation}
\begin{equation}
H_t = \frac{\hbar^2}{2 m_0} A_6 k_t k_z
\end{equation}
\begin{equation}
\Delta = \sqrt{2} \ \Delta_3
\end{equation}
\begin{equation}
\lambda = \frac{\hbar^2}{2 m_0}\left( A_1 k_z^2 +A_2 k_t^2 \right)
          +D_1 \epsilon_{zz}+ D_2 \left( \epsilon_{xx}+\epsilon_{yy} \right)
\end{equation}
\begin{equation}
\theta = \frac{\hbar^2}{2 m_0}\left( A_3 k_z^2 +A_4 k_t^2 \right)
          +D_3 \epsilon_{zz}+ D_4 \left( \epsilon_{xx}+\epsilon_{yy} \right)
\end{equation}
\end{mathletters}
In Eq. (\ref{ValenceMatrixElements}), 
the $A_i's$ are effective mass parameters, the $D_i's$ are the
Bir-Pikus deformation potentials, and the $\Delta 's$ are related to the
crystal field splitting, $\Delta_{cr}$, and spin-orbit
splitting, $\Delta_{so}$, by $\Delta_1=\Delta_{cr}$ and  
$\Delta_2=\Delta_3=\Delta_{so}/3$.
$m_0$ is the free electron mass.

\subsection{Quantized carrier states in MQW diodes}

\begin{figure}
\begin{center}
\epsfxsize=3.4in
\epsfbox{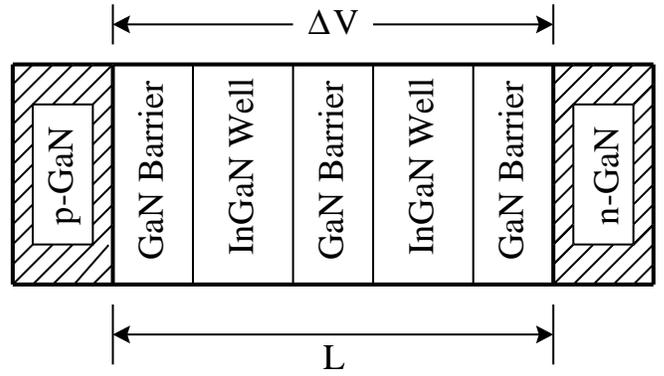}{}
\end{center}
\caption{Schematic diagram of the $In_xGa_{1-x}N$ multi quantum well diode
structure.}
\label{diode}
\end{figure}

In quantum confined systems such as the $pin$ diode shown 
in Fig.\ \ref{diode}, we must modify the bulk Hamiltonian. 
The finite MQW structure breaks translational symmetry along the $z$
direction but not in the $xy$ plane. Thus, quantum confinement of carriers
in the MQW active region gives rise to a set of two-dimensional subbands.
The wavefunctions in the envelope function approximation are
\begin{equation}
\psi^{\alpha}_{n,{\bf k}}({\bf r})= \sum_{j}\
\frac{ e^{i \ {\bf k} \cdot {\bf \rho} } }{\sqrt{A}} \ F^{\alpha}_{n, k, j}(z)
\ \arrowvert \alpha, j \rangle ,
\label{Wavefunction}
\end{equation}
where $\alpha = \{ c,v \}$ refers to conduction or valence subbands, 
$n$ is the subband index, ${\bf k} = (k_x,k_y,0) = (k,\phi)$ is the
two-dimensional wavevector, and $j$ labels the spinor component. For
conduction subbands, ($\alpha$=c)\ $j = 1,2$ \ while for valence subbands
($\alpha$=v)\ $j=1 ... 6$. The slowly varying envelope functions
$F^{\alpha}_{n, k, j}(z)$ are real and depend only on $k = |{\bf k}|$, while
the rapidly-varying Bloch basis states
$\arrowvert \alpha, j \rangle$ defined in Eqs. (\ref{cBasis}) and
(\ref{vBasis}) depend on $\phi$ in the case of valence subbands as
given in Eq. (\ref{phases}).
The area of the MQW sample in the $xy$ plane
is $A$, and ${\bf \rho} = (x,y,0)$ is the projection of {\bf r} in the plane.

The envelope functions satisfy a set of effective-mass Schr\"{o}dinger
equations
\begin{equation}
\sum_{j,j'}  \left\{ H^{\alpha}_{j,j'}( k )
+ \delta_{j,j'}\ \left[ V_{\alpha}(z)-E^{\alpha}_n( k ) \right] \right\}
F^{\alpha}_{n, k, j' }(z) =0,
\end{equation}
subject to the boundary conditions
\begin{equation}
F^{\alpha}_{n, k, j }(z=0) \ = \ F^{\alpha}_{n, k, j }(z=L) \ = \ 0.
\label{BoundaryCondition}
\end{equation}
where $L$ is again the length of the MQW diode structure (c.f. Fig. \
\ref{diode}), $V_{\alpha}(z)$ are the quantum confinement potentials
for conduction and valence electrons. $E^{\alpha}_n( k )$ are the energy
eigenvalues for the $n$th conduction or valence subband.  Note that in the
envelope function approximation, the subband energy depends only on the
magnitude $k$ of the transverse wavevector and not on the angle $\phi$. 
For the quantum confined case, the matrix 
operators $H^{\alpha}_{j,j'}(k)$ depend explicitly on $z$ and
are obtained by making the replacement
$k_{z} \rightarrow -i \frac{\partial}{\partial z}$ and letting all
material parameters be $z$-dependent operators
in the matrices $H^{\alpha}(k,\epsilon)$ given in Eqs.
(\ref{HcMatrix}) and (\ref{HvMatrix}).
To ensure the Hermitian property of the Hamiltonian, we make the
operator replacements \cite{chuang95}
\begin{mathletters}
\begin{equation}
B(z) \ \frac{\partial^2}{\partial z^2} \rightarrow 
\frac{\partial}{\partial z} \ B(z) \ \frac{\partial}{\partial z}, 
\end{equation}
and
\begin{equation}
B(z) \ \frac{\partial}{\partial z} \rightarrow
\frac{1}{2} \ \left[ B(z) \ \frac{\partial}{\partial z}
+ \frac{\partial}{\partial z} \ B(z)  \right] .
\end{equation}
\end{mathletters}

The quantum confinement potentials $V_{c}(z)$ and $V_{v}(z)$ arise from
(i) bandgap discontinuities between well and barrier regions,
(ii) the strain-induced piezoelectric field, and (iii) the time-dependent
electric field due to photoexcited electrons and holes. Thus,
\begin{equation}
V_{\alpha}(z,t) = V_{\alpha,\text{gap}}(z)+V_{\text{piezo}}(z)
+ V_{\text{photo}}(z,t) ,
\end{equation}
and the bandstructure is explicitly time-dependent.

Material parameters for $InN$ and $GaN$ used in this work can be found in
Table \ref{ParameterTable}.
For $In_{x}Ga_{1-x}N$ alloys, we interpolate between the $GaN$ and $InN$
values listed in the table.

We obtain electron effective masses by linearly
interpolating the reciprocals of the masses as a function of
the indium concentration $x$, i.e., the concentration dependent effective
masses are taken to be

\begin{mathletters}
\begin{equation}
\frac{1}{m^*_{xy}(x)} = x \left(\frac{1}{m^*_{xy}}\right)_{\text{InN}}
+ \ (1-x) \left(\frac{1}{m^*_{xy}}\right)_{\text{GaN}} ,
\end{equation}
and
\begin{equation}
\frac{1}{m^*_{z}(x)} = x \left(\frac{1}{m^*_{z}}\right)_{\text{InN}}
+ \ (1-x) \left(\frac{1}{m^*_{z}}\right)_{\text{GaN}} .
\end{equation}
\end{mathletters}

For the alloy band gap, $E_g(x)$,
we use an expression incorporating a bowing paremeter:
\begin{equation}
E_g(x)=x E_{g,InN} + (1-x) E_{g,GaN} - b \ x(1-x) , 
\label{BandGap}
\end{equation}
where the bowing parameter, $b= 1.0 \ eV$. \cite{nakamura97}

For all other material parameters, we use linear interpolation in $x$
to obtain values for the alloy. Since we can't find deformation potentials
for $InN$, we use $GaN$ values by default. In the absence of values
of $\epsilon_{\infty}$ for either $GaN$ or $InN$, we 
use linear interpolation in $x$ to obtain $\epsilon_{0}$ and simply take
$\epsilon_{\infty} \approx \epsilon_{0}$.

If the z-dependent band gap in the MQW is $E_g(z)$ as determined from
Eq.\ (\ref{BandGap}) and the Indium concentration profile,
and $E_{g,\text{min}} = \text{min}_z[E_g(z)]$ is the minimum band gap
in the structure, then the confinement potentials for conduction and 
valence electrons are defined as
\begin{mathletters}
\begin{equation}
V_{c,\text{gap}}(z)=E_{g,\text{min}}+ Q_c \ ( E_g(z)-E_{g,\text{min}} )
\end{equation}
\begin{equation}
V_{v,\text{gap}}(z)=- (1-Q_c)\ ( E_g(z)-E_{g,\text{min}} ) 
\end{equation}
\end{mathletters}
where the conduction band offset is taken as $Q_c=0.6$. \cite{nakamura97}
With these definitions for the $V_{\alpha,\text{gap}}(z)$, the zero of the
gap confinement potential is placed at the top of the valence band profile.

The confinement potentials due to the strain-induced piezoelectric field
are given by
\begin{equation}
V_{\text{piezo}}(z) = -|e| \ E^0_z(z) ,
\end{equation}
where $|e|$ is the electric charge and $E^0_z(z)$ is the
strain-induced piezoelectric field.
In a pseudomorphically strained MQW diode, the bulk 
source and drain (assumed to have identical composition) 
are unstrained while the in-plane MQW lattice
constants adjust to the source and drain values.
For a MQW grown along [0001] (the z-direction) the z-dependent strain
is \cite{wright97}
\begin{equation}
\epsilon_{xx}(z)=\epsilon_{yy}(z)=\frac{a_0-a(z)}{a(z)}.
\label{Epsilonxx}
\end{equation}
Here $a_0$ is the lattice constant in the source and drain and $a(z)$ is
the z-dependent lattice constant in the MQW. Minimizing the
overall strain energy, we find \cite{wright97}
\begin{equation}
\epsilon_{zz}(z)=-\ \frac{2 \ C_{13}(z)}{C_{33}(z)} \ \epsilon_{xx}(z) ,
\label{Epsilonzz}
\end{equation}
where $C_{13}(z)$ and $C_{33}(z)$ are z-dependent elastic constants.

There are several issues concerning strain that one can worry about. One
is the critical well thickness beyond which the strain relaxes. Studies have
shown \cite{chichibu99} that with 10 \% Indium, and 6 nm well width, a 350 kV/cm
field is measured, implying that the well is not fully relaxed and justifies
using the pseudomorphic strain approximation as we do in this paper
for wells having 6 \% Indium and thickness near 4 nm. For thick
wells, this pseudormorphic strain approximation clearly will begin to break down
and a more detailed model will be needed. Interface roughness can also play a
role. The roughness will not significantly affect the acoustic phonon modes
(see introduction) but may affect the photogeneration of carriers \cite{note1}.
For simplicity, we do not consider interface roughness.

The strain-induced polarization directed along the z-direction
is given by
\begin{equation}
P_z^0(z)=e_{31}(z)\ \left(\epsilon_{xx}(z)+\epsilon_{yy}(z) \right)
+e_{33}(z)\ \epsilon_{zz}(z) ,
\label{PzEq}
\end{equation}
where $e_{31}(z)$ and $e_{33}(z)$ are z-dependent piezoelectric
constants. The unscreened piezoelectric field in the diode is obtained from
the requirement that the electric displacement vanishes \cite{smith88}. Thus,
\begin{equation}
E_z^0(z)=-\ \frac{4 \pi}{\varepsilon_0(z)} \
\left( P_z^0(z) + P_0 \right) ,
\label{EzEq}
\end{equation}
where $P_0$ is a constant polarization induced by externally applied
voltages and $\varepsilon_0(z)$ is the position-dependent static dielectric
constant. The value of $P_0$ is obtained from the voltage
drop across the diode (of length $L$) in the unscreened limit, i.e. with no
photoexcited carriers.
In this limit, the voltage drop between source and drain due to the induced
piezoelectric field is just
\begin{equation}
V_A = - \int_{0}^{L} dz \ E_z^0(z) ,
\label{P0Eq}
\end{equation}
from which $P_0$ can be determined.

When  photoexcited electrons and holes are generated by the laser, then
there is an additional time-dependent confinement potential,
\begin{equation}
V_{\text{photo}}(z,t)=-|e| \ E_z^{\text{photo}}(z,t).
\end{equation}
This potential is obtained by solving the
Poisson equation in the diode for $E_z^{\text{photo}}(z,t)$ subject
to the boundary condition
\begin{equation}
V_A = - \int_{0}^{L} dz \ E_z^{\text{total}}(z,t) .
\end{equation}
Here, $ E_z^{\text{total}}(z,t)$ is the total electric field 
and is just the sum of the strain induced
electric field and the field due to photogenerated electrons and holes, i.e.,
\begin{equation}
E_z^{\text{total}}(z,t)=E_z^0(z) + E_z^{\text{photo}}(z,t) .
\end{equation}

Finally, we can write an 
effective-mass Schr\"{o}dinger equation for the conduction electron envelope
functions in terms of an effedctive electron potential, $ V^{eff}_c(z)$:
\begin{eqnarray}
&& -\frac{\hbar^2}{2} \ 
\left\{  \frac{\partial}{\partial z}\ \frac{1}{m^{*}_{z}(z)}\ 
\frac{\partial}{\partial z} \right\} \ F^{c}_{n, k, j}(z)
\nonumber \\
&& + \left\{ V^{eff}_c(z)- E^c_n( k ) \right\} \ F^{c}_{n, k, j}(z)=0 , 
\end{eqnarray}
where the effective electron potential is
\begin{eqnarray}
&& V^{eff}_c(z)\ = V_c(z) + \frac{\hbar^2 k^2}{2 \  m^{*}_{xy}(z)}
\nonumber \\
&& + \ a_{c,z}(z) \epsilon_{zz}(z)+a_{c,xy}(z) \
\left( \epsilon_{xx}(z)+\epsilon_{yy}(z) \right).
\end{eqnarray}
Similar expressions can be derived for valence electrons. We arrive
at a set of coupled ordinary differential equations (ODE's) subject
to the two-point boundary value condition of
Eq.\ (\ref{BoundaryCondition}).
These are solved for the envelope functions and subband energies.

In practice, we introduce a uniform grid, $\{ z_i \}$, along the z-direction 
and finite-difference the effective mass Schr\"{o}dinger equations
to obtain a matrix eigenvalue problem which can be solved using
standard matrix eigenvalue routines. The resulting eigenvalues are the subband
energies, $E^{\alpha}_n( k )$, and the corresponding eigenvectors are
the envelope functions, $F^{\alpha}_{n,k,j}(z_i)$, defined on the finite
difference mesh.

\subsection{Second quantized electron hamiltonians}


We next describe the second quantized Hamiltonians for electrons
moving freely in the MQW interacting via a screened Coulomb potential.
We denote creation and destruction operators for electrons in conduction
and valence subbands by $c^{\dagger}_{\alpha,n,{\bf k}}$ and
$c_{\alpha,n,{\bf k}}$, respectively.
The second quantized Hamiltonian for free electrons and holes is simply
\begin{equation}
{\cal{H}}_{e0}=\sum_{\alpha,n,{\bf k}} \ E^\alpha_n(k) \
 c^{\dagger}_{\alpha,n,{\bf k}} c_{\alpha,n,{\bf k}} .
\end{equation}


The Coulomb interaction Hamiltonian is given by
\begin{eqnarray}
{\cal{H}}_{ee} &=& \frac{1}{2}\sum_{\alpha,n,{\bf k}} \
\sum_{\alpha',n',{\bf k}'} \sum_{{\bf \kappa} \neq 0}
\ {\cal{V}}^{\alpha,n, \ {\bf k}}_{\alpha',n',{\bf k}'}({\bf \kappa})
\nonumber \\
&& \times \ c^{\dagger}_{\alpha,n,{\bf k}-{\bf \kappa}}
c^{\dagger}_{\alpha',n',{\bf k}'+{\bf \kappa}}
\ c_{\alpha',n',{\bf k}'} c_{\alpha,n,{\bf k}} .
\label{CoulombHamiltonian}
\end{eqnarray}
Eq.\ (\ref{CoulombHamiltonian}) describes two-body interactions where
electrons in states $\arrowvert \alpha, n, {\bf k} \rangle$
and $\arrowvert \alpha', n', {\bf k}' \rangle$ scatter to subband states
$\arrowvert \alpha, n, {\bf k}-{\bf \kappa} \rangle$ and 
$\arrowvert \alpha', n', {\bf k}'+{\bf \kappa} \rangle$, respectively.
Note that to simplify things,
we have neglected terms corresponding to Coulomb induced interband transitions
(the "diagonal approximation")
since these are very unfavorable energetically. \cite{chow94} The electrons
thus stay in their original subbands (though they may scatter off
of other electrons in different subbands) and exchange crystal momentum
${\bf \kappa}$.
The matrix elements describing the strength of these transitions are
given by 
\begin{eqnarray}
{\cal{V}}^{\alpha,n, \ {\bf k}}_{\alpha',n',{\bf k}'}({\bf \kappa})\ &=&
\int dz \int dz' \ V_{|{\bf \kappa}|}(z-z')
\nonumber \\
&& \times \ \sum_{j}
F^{\alpha}_{n, |{\bf k}-{\bf \kappa}|, j }(z)\ F^{\alpha}_{n,k,j }(z)
\nonumber \\
&& \times \ \sum_{j'}
F^{\alpha'}_{n',|{\bf k}'+{\bf \kappa}|,j}(z')\ F^{\alpha'}_{n',k',j'}(z') \ .
\label{ExactVMatrix}
\end{eqnarray}
From Eq.\ (\ref{ExactVMatrix}), it is apparent that the symmetry relation
\begin{equation}
{\cal{V}}^{\alpha,n, \ {\bf k}}_{\alpha',n',{\bf k}'}({\bf \kappa}) =
{\cal{V}}^{\alpha',n',{\bf k}'}_{\alpha,n,\ {\bf k}}({\bf -\kappa})
\label{Symmetry}
\end{equation}
must hold.

The Fourier transform in the $xy$ plane of the screened Coulomb
potential depends only on $\kappa \equiv | {\bf \kappa} |$ and $|z|$
and is given by
\begin{equation}
V_{\kappa}(z)=\frac{2\pi e^2}{\epsilon_0 A}
\ \frac{ e^{-\kappa |z|} }{ \kappa \ \epsilon_s( \kappa ) } \ .
\end{equation}
To describe screening, we adopt an effective pseudodynamic
dielectric function of the form
\begin{equation}
\frac{1}{\epsilon_s(\kappa)}=\frac{\kappa}{\kappa+\kappa_s} \ .
\end{equation}
In the pseudodynamic screening model, we completely neglect screening
by the massive holes and treat screening by the lighter
conduction electrons in the static screening limit. 
The screening wavevector, $\kappa_s$, is computed in the $2D$ limit.
Thus \cite{haug93} 
\begin{equation}
\kappa_s=\frac{2\pi e^2}{\epsilon_0}\ \frac{\partial N_{2D}}{\partial \mu} \ , 
\end{equation}
where $N_{2D}$, the two-dimensional conduction electron density, is
related to an effective chemical potential, $\mu$, by
\begin{equation}
N_{2D} = \frac{m^{*}_{xy} \ k_B T}{2\pi \ \hbar^2}
\sum_n \ln \left[
1+\exp \left(  \frac{E^c_{n}(0) - \mu}{k_B T} \right)
\right] \ .
\label{N2d}
\end{equation}
In Eq.\ (\ref{N2d}), $E^c_n(0)$ is the conduction subband energy evaluated
at $k = 0$.
In our simulation,
the value of $\mu$ is obtained by requiring that $N_{2D}$, evaluated using
Eq.\ (\ref{N2d}), be equal to the value
\begin{equation}
N_{2D}(t) = \sum_{n, {\bf k} } f^c_n( k,t )
\label{N2dt}
\end{equation}
obtained from the time dependent conduction electron distribution
functions, $f^c_n(k,t)$.

\subsection{Photogeneration of carriers}


Electron-hole pairs are created by the pump laser and we treat the electric
field of the laser in the semiclassical dipole approximation. 
In this approximation, the electron-laser interaction Hamiltonian is
\begin{equation}
{\cal{H}}_{eL}=-\arrowvert e \arrowvert \
{\bf E(t)} \cdot \sum_{n,n',{\bf k}}
\left[
{\bf d}^{c,v}_{n,n'}({\bf k}) c^{\dagger}_{c,n,{\bf k}}c_{v,n',{\bf k}}
+ \text{h.c.}
\right] ,
\end{equation}
where h.c. denotes the Hermitian conjugate of the first term. The laser
field is ${\bf E(t)}$ and the dipole matrix elements are
\begin{equation}
{\bf d}^{c,v}_{n,n'}({\bf k})
=\sum_{j,j'}{\bf D}^{c,v}_{j,j'}(\phi)
\ \int dz \ F^c_{n,k,j}(z)\ F^v_{n',k,j'}(z) \ . 
\label{dcv}
\end{equation}
The vector operator, ${\bf D}^{c,v}_{j,j'}(\phi)$, is a $2 \times 6$
matrix with $x$, $y$, and $z$ components. Thus,
\begin{equation}
{\bf D}^{c,v}_{j,j'}(\phi) \equiv D^{c,v}_{X}(\phi) \ {\bf \hat{x}} +
D^{c,v}_{Y}(\phi) \ {\bf \hat{y}} + D^{c,v}_{Z}(\phi) \ {\bf \hat{z}} \ ,
\end{equation}
where ${\bf \hat{x}}$, ${\bf \hat{y}}$, and ${\bf \hat{z}}$ are unit
vectors and
\begin{mathletters}
\label{Dcv}
\begin{equation}
D^{c,v}_{X}(\phi) = \frac{P_2}{\sqrt{2} \ E_g}
\left[
\begin{array}{cccccc}
 \alpha   & -\beta^* & 0 &  \alpha   & -\beta^* & 0 \\
-\alpha^* &  \beta   & 0 &  \alpha^* & -\beta   & 0
\end{array}
\right] \ .
\end{equation}
\begin{equation}
D^{c,v}_{Y}(\phi) = \frac{i \ P_2}{\sqrt{2} \ E_g}
\left[
\begin{array}{cccccc}
-\alpha   & -\beta^* & 0 & -\alpha   & -\beta^* & 0 \\
-\alpha^* & -\beta   & 0 &  \alpha^* &  \beta   & 0
\end{array}
\right] ,
\end{equation}
\begin{equation}
D^{c,v}_{Z}(\phi) = \frac{P_1}{E_g}
\left[
\begin{array}{cccccc}
0 & 0 & -\beta   & 0 & 0 &  \beta \\
0 & 0 & -\beta^* & 0 & 0 & -\beta^*
\end{array}
\right] .
\end{equation}
\end{mathletters}
The $6 \times 2 $ vector operator, ${\bf D}^{v,c}_{j',j}(\phi)$, is related
to the $2 \times 6$ operator, ${\bf D}^{c,v}_{j,j'}(\phi)$, by
\begin{equation}
{\bf D}^{v,c}_{j',j}(\phi) = \left( {\bf D}^{c,v}_{j,j'}(\phi) \right)^* \ .
\end{equation}

In Eq.\ (\ref{Dcv}),
$\alpha$ and $\beta$ are the $\phi$-dependent phase factors defined
in Eq.\ (\ref{phases}) and
the Kane parameters, $P_1$ and $P_2$, for wurtzite materials are
related to the effective masses and energy gaps by \cite{park99}
\begin{mathletters}
\begin{eqnarray}
P_1^2 &=& \frac{\hbar^2}{2 m_0} \left( \frac{m_0}{m^*_{z}}-1 \right)
\nonumber \\
&\times& \frac{(E_g+\Delta_1+\Delta_2)(E_g+2 \Delta_2)-2\Delta_3^2}
{E_g+2 \Delta_2} ,
\end{eqnarray}
\begin{eqnarray}
P_2^2 &=& \frac{\hbar^2}{2 m_0} \left( \frac{m_0}{m^*_{xy}}-1 \right)
\nonumber \\
&\times& \frac{E_g\{(E_g+\Delta_1+\Delta_2)(E_g+2 \Delta_2)-2\Delta_3^2\}}
{(E_g+\Delta_1+\Delta_2)(E_g+\Delta_2)-\Delta_3^2} \ .
\end{eqnarray}
\end{mathletters}

For the semiclassical laser field, we write the real electric
field as
\begin{equation}
{\bf E}(t) =
\frac{1}{2} \left[ \ {\bf \hat{\epsilon}} \ {\cal{E}}(t) \ e^{i  \omega t}
+ {\bf \hat{\epsilon}}^{*} \ {\cal{E}}(t) \ e^{-i \omega t} \ \right] \,
\end{equation}
where $\omega$ is the photon frequency, ${\bf \hat{\epsilon}}$ is
a complex unit polarization vector, and ${\cal{E}}(t)$ is the pulse shape
envelope function. We assume a Gaussian pulse shape
\begin{equation}
{\cal{E}}(t) = {\cal{E}}_0 \ \exp \left[
- \left( \frac{t-t_0}{\tau \sqrt{\frac{1}{2 \ln 2}}} \right)^2 \right]
\label{GaussianPulseShape}
\end{equation}
centered at $t = t_0$ with an intensity full width at half maximum (FWHM)
of $\tau$. The maximum electric field strength, ${\cal{E}}_0$, is
related to the pump fluence, ${\cal{F}}$, by
\begin{equation}
{\cal{E}}_0 = \sqrt{ \frac{16\pi \ {\cal{F}}}{c\ n_\omega \tau}
\sqrt{\frac{\ln 2}{\pi}} } \ ,
\end{equation}
where $n_\omega$ is the index of refraction at the photon frequency.

For linearly polarized light incident normally on the MQW, the polarization
vectors are real and given by either ${\bf \hat{x}}$ or ${\bf \hat{y}}$.
For circularly polarized light, the polarization vectors are complex and
given by \cite{jackson75}
\begin{equation}
{\bf\hat{\epsilon}}_{\pm}=\frac{{\bf\hat{x}}\ \pm \ i{\bf\hat{y}}}{\sqrt{2}}
\label{polarization} \ .
\end{equation}
In Eq.\ (\ref{polarization}), the upper sign refers to left circularly
polarized light (positive helicity) and the lower sign refers to right
circularly polarized light (negative helicity).

\subsection{Coupling to LA phonons}


We treat the acoustic phonons in the MQW as bulk-like plane-wave
states with wavevector ${\bf q}$. Since the system exhibits cylindrical
symmetry, only ${\bf q} = q\ {\bf \hat{z}}$ 
longitudinal acoustic phonons are coupled by the electron-phonon interaction.
The free LA phonon Hamiltonian can be written as
\begin{equation}
{\cal{H}}_{A0}= \sum_{q} \ \hbar \omega_q \ b^{\dagger}_q b_q \ .
\end{equation}
where $b^{\dagger}_q$ and $b_q$ are creation and destruction
operators for LA phonons with wavevector ${\bf q} = q\ {\bf \hat{z}}$.
The wave vector component, $q$, of LA phonons in the MQW is thus defined in
an extended zone scheme where $-\infty < q < \infty$. The phonon
dispersion relation is given by a linear relation
\begin{equation}
\omega_q = C_s\ |q| = \sqrt{ \frac{C_{33}}{\rho_0} }\ |q| \ , 
\label{PhononDispersion}
\end{equation}
where $\rho_0$ is the mass density and $C_s$ is just the LA phonon sound speed
for propagation parallel to ${\bf \hat{z}}$. \cite{yu96}
In computing the LA sound speed in the linear phonon dispersion relation
of Eq. (\ref{PhononDispersion}),
we neglect the $z$-dependence of the material parmeters and use bulk $GaN$
values for $C_{33}$ and $\rho_0$.


The LA phonons in wurtzite MQW's interact with the electrons through
deformation potential and screened piezoelectric scattering.
The electron-LA phonon interaction in an MQW is governed by the Hamiltonian
\begin{equation}
{\cal{H}}_{eA}= \sum_{\alpha,n,n',{\bf k},q}
{\cal{M}}^{\alpha}_{n,n'}(k,q) (b_q + b^\dagger_{-q})
\ c^{\dagger}_{\alpha,n,{\bf k}}c_{\alpha,n',{\bf k}} \ .
\end{equation}
This Hamiltonian describes the scattering of an electron from subband
state $ \arrowvert \alpha, n', {\bf k} \rangle $
to subband state $\arrowvert \alpha, n, {\bf k} \rangle$ with 
either the emission or absorption of an LA phonon. We note that the
electron wave vector, $\bf k$, in the $xy$ plane is conserved in this process
since, as noted earlier, the phonon wave vector in the $xy$ plane is zero.

The interaction matrix elements describing deformation and screened
piezoelectric scattering are 
\begin{eqnarray}
{\cal{M}}^{\alpha}_{n,n'}&&(k,q)=
\sqrt{\frac{\hbar^2}{2 \rho_0 \ (\hbar \omega_q)\ V}}
\nonumber \\
&&\times \left[
iq \ {\cal{D}}^{\alpha}_{n,n'}(k,q)
-\frac{4 \pi |e| \ e_{33}}{\epsilon_{\infty}\ \epsilon_s(q)}\
{\cal{P}}^{\alpha}_{n,n'}(k,q)
\right] \ ,
\label{PhononMatrix}
\end{eqnarray}
where $V$ is the crystal volume.
The first term in Eq.\ (\ref{PhononMatrix}) desribes deformation potential
scattering while the second term describes screened piezoelectric scattering.

The relative strengths of the various transitions are determined
by form factors for deformation potential and piezoelectric scattering.
The form factor for screened piezoelectric scattering is given by
\begin{equation}
{\cal{P}}^{\alpha}_{n,n'}(k,q)= \sum_{j}
\int dz \ F^{\alpha}_{n,k,j}(z)\ e^{iqz}\ F^{\alpha}_{n',k,j}(z) \ ,
\end{equation}
while the form factor for deformation potential scattering is
defined to be
\begin{equation}
{\cal{D}}^{\alpha}_{n,n'}(k,q)= \sum_{j} \Theta^{\alpha}_{j}
\int dz \ F^{\alpha}_{n,k,j}(z)\ e^{iqz}\ F^{\alpha}_{n',k,j}(z) \ . 
\end{equation}

The form factor for deformation potential scattering 
is similar to the form factor for piezoelectric scattering except that
in summing over spinor components, $j$, the terms are weighted
by $j$-dependent deformation potentials $\Theta^{\alpha}_{j}$ which 
can be representd by the row vectors
\begin{mathletters}
\begin{equation}
\Theta^c_j=\left\{ \ a_{c,z}, \ a_{c,z}\ \right\} ,
\end{equation} 
\begin{equation}
\Theta^v_j=\left\{ D_1+D_3, D_1+D_3, D_1, D_1+D_3, D_1+D_3, D_1 \right\}.
\end{equation}
\end{mathletters}
for conduction and valence electrons, respectively.

\subsection{Electron density matrices}

We define statistical operators in terms of the electron and phonon
eigenstates. The electron density matrix is
\begin{equation}
N^{\alpha,\alpha'}_{n,n'}({\bf k},t) \equiv \left< \
c^{\dagger}_{\alpha,n,{\bf k}}(t)\ c_{\alpha',n',{\bf k}}(t) \ \right> ,
\end{equation}
where $\langle \ \rangle$ denotes the statistial average of the
non-equlilbrium state of the system.

The interband components of the density matrix,
$N^{c,v}_{n,n'}({\bf k},t)$ and $N^{v,c}_{n',n}({\bf k},t)$,
describe the coherence between conduction and
valence electrons in subbands $n$ and $n'$ and are related to the
optical polarization. The intraband components of the density matrix,
$N^{\alpha,\alpha}_{n,n'}({\bf k},t)$ describe correlations between
different subbands of the same carrier type if $n \neq n'$. If $n=n'$,
$N^{\alpha,\alpha}_{n,n}({\bf k},t) \equiv f^{\alpha}_{n}({\bf k},t)$ is just
the carrier distribution function for electrons in the subband state,
$\psi^{\alpha}_{n,{\bf k}}({\bf r})$, defined in Eq.\ (\ref{Wavefunction}).

\subsection{Coherent phonon amplitude}

The coherent phonon amplitude of the $q$-th phonon mode,
$\arrowvert q \rangle$, is defined to be \cite{kuznetsov94}
\begin{equation}
D_q(t) \equiv \left< \ b_{q}(t) + b^{\dagger}_{-q}(t) \ \right> \ .
\end{equation}
The coherent phonon amplitude is related to the macroscopic lattice
displacement, $U(z,t)$, and velocity, $V(z,t)$, through the relations
\begin{equation}
U(z,t)=\sum_q \sqrt{\frac{\hbar^2}{2 \rho_0 \ (\hbar \omega_q) \ V}}
\ e^{i q z} \ D_q(t)
\label{Uzt}
\end{equation}
\begin{equation}
V(z,t)=\sum_q \sqrt{\frac{\hbar^2}{2 \rho_0 \ (\hbar \omega_q) \ V}}
\ e^{i q z} \ \frac{\partial D_q(t)}{\partial t}
\label{Vzt}
\end{equation}

The coherent phonon amplitude, $D_q(t)$, will vanish if there are a
definite number of phonons in the mode, i.e., if the phonon oscillator
is in one of its energy eigenstates, $\arrowvert q \rangle$. In this
case, there is no macroscopic displacement of the lattice.

The coherent phonon distribution is \cite{kuznetsov94}
\begin{equation}
{\cal{N}}^{coh}_q(t)
\equiv \langle b^{\dagger}_q(t) \rangle \ \langle b_q(t) \rangle
\end{equation}
and the total phonon distribution, ${\cal{N}}_q(t)$, can be separated into
coherent and incoherent contributions as follows;
\begin{eqnarray}
{\cal{N}}_q(t) && = \langle  b_q(t)\ b^{\dagger}_{-q}(t)\ \rangle \
\nonumber \\
&& \equiv {\cal{N}}^{coh}_q(t) + {\cal{N}}^{incoh}_q(t) \ .
\end{eqnarray}
In general, a mode can have a number of both coherent and incoherent 
phonons, but only the coherent phonons contribute to the macroscopic
lattice displacement.

We note that at the beginning of the experiment, there are no coherent
phonons present, i.e. ${\cal{N}}^{coh}_q(t)=0$, and the incoherent phonon
population is described by a thermal distribution,
${\cal{N}}^{incoh}_q(t) \sim e^{-\hbar \omega_q /k_B T }$.

\subsection{Equations of motion}

In this section we develop equations of motion for the electron density
matrices and coherent phonon amplitudes. The electron density matrices
obey the general equations of motion

\begin{equation}
\frac{\partial N^{\alpha,\alpha'}_{n,n'}({\bf k},t)}{\partial t} =
\left<
\frac{i}{\hbar}
\left[ {\cal{H}},\ c^{\dagger}_{\alpha,n,{\bf k}}c_{\alpha',n',{\bf k}} \right]
\right> \ ,
\label{BlochEquation}
\end{equation}
where $[ \ ]$ denotes the commutator and $\langle \ \rangle$ denotes the
average over an initial ensemble. The density matrices are defined in the
electron picture and initially the valence bands are filled while the
conduction bands are empty. We have $f^c_n({\bf k},t=-\infty)=0$ and
$f^v_n({\bf k},t=-\infty)=1$ which implies
\begin{equation}
N^{\alpha,\alpha'}_{n,n'}({\bf k},t=-\infty)=
\delta_{n,n'} \  \delta_{\alpha,v} \ \delta_{\alpha',v} \ .
\end{equation}
The total Hamiltonian, ${\cal{H}}$, is the sum of the Hamiltonians
described in the previous sections, i.e. 
\begin{equation}
{\cal{H}}={\cal{H}}_{e0} + {\cal{H}}_{ee} + {\cal{H}}_{eL} +
{\cal{H}}_{A0} + {\cal{H}}_{eA} \ .
\end{equation}

In deriving equations of motion for the density matrices, we make the
ansatz that the density matrices depend only on $k=|{\bf k}|$.
We use the rotating wave approximation (RWA) to factor out the rapid
$e^{i \omega t}$ behavior of the interband density matrix elements,
$N^{c,v}_{n,n'}(k,t)$. In the RWA, we have
\begin{equation}
N^{c,v}_{n,n'}(k,t) \equiv \tilde{N}^{c,v}_{n,n'}(k,t) \ e^{i \omega t} \ ,
\end{equation}
where $\tilde{N}^{c,v}_{n,n'}(k,t)$ is a slowly varying envelope function.
In addition, we treat the Coulomb interaction in the time-dependent
Hartree-Fock approximation by factoring four-operator averages arising
from ${\cal{H}}_{eL}$ into appropriate products of two-operator averages
as described in Ref.\ \onlinecite{haug93}. 

The resulting equations of motion for the density matrices are
\begin{mathletters}
\label{BlochEquations}
\begin{eqnarray}
\frac{\partial N^{c,c}_{n,n'}(k,t)}{\partial t} =
\frac{i}{\hbar} \left\{ 
{\cal{E}}^{c}_{n}(k)-{\cal{E}}^{c}_{n'}(k)
\right\} N^{c,c}_{n,n'}(k)
\nonumber \\
-i \sum_m
\left\{
 \Omega^{c,v}_{n,m}(k) \tilde{N}^{v,c}_{m,n'}(k)
-\tilde{N}^{c,v}_{n,m}(k) \Omega^{v,c}_{n,n'}(k)
\right\}
\nonumber \\
+ \frac{i}{\hbar} {\sum_{m}}'
\left\{
\Lambda^{c}_{n,m}(k)  N^{c,c}_{m,n'}(k)
-N^{c,c}_{n,m}(k) \Lambda^{c}_{m,n'}(k)
\right\}
\end{eqnarray}
\begin{eqnarray}
\frac{\partial N^{v,v}_{n,n'}(k,t)}{\partial t} =
\frac{i}{\hbar} \left\{ 
{\cal{E}}^{v}_{n}(k)-{\cal{E}}^{v}_{n'}(k)
\right\} N^{v,v}_{n,n'}(k)
\nonumber \\
-i \sum_m
\left\{
 \Omega^{v,c}_{n,m}(k) \tilde{N}^{c,v}_{m,n'}(k)
-\tilde{N}^{v,c}_{n,m}(k) \Omega^{c,v}_{n,n'}(k)
\right\}
\nonumber \\
+ \frac{i}{\hbar} {\sum_{m}}'
\left\{
\Lambda^{v}_{n,m}(k)  N^{v,v}_{m,n'}(k)
-N^{v,v}_{n,m}(k) \Lambda^{v}_{m,n'}(k)
\right\}
\end{eqnarray}
\begin{eqnarray}
\frac{\partial \tilde{N}^{c,v}_{n,n'}(k,t)}{\partial t} =
\frac{i}{\hbar} \left\{ 
{\cal{E}}^{c}_{n}(k)-{\cal{E}}^{v}_{n'}(k) - \hbar \omega
\right\} \tilde{N}^{c,v}_{n,n'}(k)
\nonumber \\
- i \sum_m
\left\{
 \Omega^{c,v}_{n,m}(k) N^{v,v}_{m,n'}(k)
-N^{c,c}_{n,m}(k) \Omega^{c,v}_{m,n'}(k)
\right\}
\nonumber \\
+ \frac{i}{\hbar} {\sum_{m}}'
\left\{
\Lambda^{c}_{n,m}(k)  \tilde{N}^{c,v}_{m,n'}(k)
-\tilde{N}^{c,v}_{n,m}(k) \Lambda^{v}_{m,n'}(k)
\right\} \ .
\end{eqnarray}
\end{mathletters}
The equations of motion for $\tilde{N}^{v,c}_{n,n'}(k,t)$ are redundant since
$\tilde{N}^{v,c}_{n,n'}(k,t)=(\tilde{N}^{c,v}_{n',n}(k,t))^{*}$.

The first terms on the right-hand side of Eq.\ (\ref{BlochEquations})
describe the free oscillation of the density matrices
in the renormalized singel particle energy bands.
The time-dependent single particle energies are
\begin{equation}
{\cal{E}}^{\alpha}_{n}(k,t)=E^{\alpha}_{n}(k)+\Lambda^{\alpha}_{n,n}(k,t) \ , 
\label{HartreeEnergies}
\end{equation}
where $E^{\alpha}_{n}(k)$ are the single particle subband energies in
the absence of conduction electrons and holes and
$\Lambda^{\alpha}_{n,n}(k,t)$ describes the time-dependent renormalization
of the single particle subbands.

The renormalization energies, $\Lambda^{\alpha}_{n,n}(k,t)$, are the 
diagonal elements of a generalized renormalization energy matrix
(in the subband indices)
\begin{equation}
\Lambda^{\alpha}_{n,n'}(k,t) =
\Sigma^{\alpha}_{n,n'}(k,t) + Q^{\alpha}_{n,n'}(k,t) \ . 
\label{Renormalization}
\end{equation}
The first term in the renormalization energy matrix (\ref{Renormalization}) 
is the generalized exchange self-energy matrix arising from the Coulomb
interaction and is given by
\begin{eqnarray}
\label{SelfEnergies}
\Sigma^{\alpha}_{n,n'}(k,t) \equiv -\sum_{{\bf k}' \neq {\bf k}}
{\cal{V}}^{\alpha, n , k}_{\alpha,n',k'}
\left( \ \arrowvert {\bf k} - {\bf k}' \arrowvert \ \right)
\nonumber \\ \times
\left( N^{\alpha,\alpha}_{n,n'}(k',t) -
\delta_{\alpha,v} \delta_{n,n'}  \right) \ ,
\end{eqnarray}
where ${\cal{V}}^{\alpha, n , k}_{\alpha,n',k'}
( \ \arrowvert {\bf k} - {\bf k}' \arrowvert \ )$ are angular averaged
Coulomb interaction matrix elements. The second term in
Eq.\ (\ref{Renormalization}) accounts for renormalization due to
coupling of carriers to coherent acoustic phonons. We have 
\begin{equation}
Q^{\alpha}_{n,n'}(k,t) \equiv \sum_q D_q(t)\ {\cal{M}}^{\alpha}_{n,n'}(k,q) , 
\end{equation}
where $D_q(t)$ is the coherent phonon amplitude and
the electron-phonon matrix elements, ${\cal{M}}^{\alpha}_{n,n'}(k,q)$,
are defined in Eq.\ (\ref{PhononMatrix}).  The self energy corrections in 
Eq.\ (\ref{SelfEnergies}) are small, though the can be important in some 
circumstances.

In computing the angular averaged 
Coulomb matrix elements in Eq.\ (\ref{SelfEnergies}), we assume
small momentum transfer, ${\bf \kappa}$, and use
the fact that the envelope functions, $F^{\alpha}_{n,k,j }(z)$, depend
weakly on $k$ to obtain an effective interaction,
\begin{eqnarray}
{\cal{V}}^{\alpha,n,k}_{\alpha',n',k'}(\kappa)\
\equiv \int dz \int dz' \ V_{\kappa}(z-z')
\nonumber \\
\times \ \sum_{j,j'}
\left< \arrowvert F^{\alpha }_{n, (k,k'), j }(z ) \arrowvert^2 \right> \
\left< \arrowvert F^{\alpha'}_{n',(k,k'), j'}(z') \arrowvert^2 \right>
\label{Vapprox}
\end{eqnarray}
where, by definition,
\begin{equation}
\left< \arrowvert F^{\alpha }_{n, (k,k'), j }(z ) \arrowvert^2 \right>
\equiv \frac{
\arrowvert F^{\alpha }_{n, k, j }(z ) \arrowvert^2 +
\arrowvert F^{\alpha }_{n, k', j }(z ) \arrowvert^2 }{2} .
\end{equation}
The effective Coulomb interaction, ${\cal{V}}$, defined
in Eq.\ (\ref{Vapprox}) is an even function of
$\kappa$ and is symmetric in $\alpha$ and $k$, thus preserving the symmetry
relation (\ref{Symmetry}). Preserving this symmetry is essential in order to
maintain conservation of carriers in the scattering process.

The second terms in Eq.\ (\ref{BlochEquations}) describe photoexcitation of
electron-hole pairs by the pump laser. The system reacts to an effective
field which is the sum of the applied field and the dipole field of the
electron-hole excitations. This gives rise to a matrix of generalized Rabi
frequencies in the subband indices
\begin{eqnarray}
\hbar \Omega^{c,v}_{n,n'}(k) = \frac{ {\cal{E}}(t) }{2} \ d^{c,v}_{n,n'}(k) +
\\ \nonumber
\sum_{{\bf k}' \neq {\bf k}}
{\cal{V}}^{c, n , k}_{v,n',k'}
\left( \ \arrowvert {\bf k} - {\bf k}' \arrowvert \ \right)
 \tilde{N}^{c,v}_{n,n'}(k',t) ,
\end{eqnarray}
which can be shown to satisfy the symmetry relations
\begin{equation}
\hbar \Omega^{c,v}_{n,n'}(k) = \left( \hbar \Omega^{v,c}_{n',n}(k) \right)^{*}
\end{equation}

The Gaussian pump envelope function, ${\cal{E}}(t)$, is defined in
Eq. (\ref{GaussianPulseShape}) and the optical dipole matrix elements 
\begin{equation}
d^{c,v}_{n,n'}(k) = \left( d^{v,c}_{n',n}(k) \right)^{*}
\equiv \int_{-\pi}^{\pi}
d \phi \ {\bf \hat{\epsilon}} \cdot {\bf d}^{c,v}_{n,n'}({\bf k})
\end{equation}
are angular averages in the $xy$ plane of the vector dipole matrices
dotted into the polarization vector. From Eq.\ (\ref{dcv}),
the $\phi$-dependence of ${\bf d}^{c,v}_{n,n'}({\bf k})$ only appears in
${\bf D}^{c,v}_{j,j'}(\phi)$ and we can get the angular averages by setting
$\alpha(\phi)=\alpha_{avg}=(1-i)/3 \pi$ and
$\beta(\phi)=\beta_{avg}=(1+i)/\pi$ in the $2 \times 6$ matrices
$D^{c,v}_{X}(\phi)$, $D^{c,v}_{Y}(\phi)$ and $D^{c,v}_{Z}(\phi)$
defined in Eq.\ (\ref{Dcv}).

The last terms in Eq.\ (\ref{BlochEquations}) are similar
in structure to the renormalization corrections in the Hartree-Fock
energies but are more complicated due to mixing among subbands and involve the
off-diagonal components of $\Lambda^{\alpha}_{n,n'}$. The prime on the
summation sign indicates that terms containing factors of
$N^{\alpha,\alpha'}_{n,n'}(k)$ are excluded from the sum since these terms
have already been incorporated in the renormalized Hartree-Fock energies
in Eq.\ (\ref{HartreeEnergies}).

The coherent phonon amplitudes, $D_q(t)$, satisfy the driven harmonic
oscillator equations
\begin{eqnarray}
\frac{\partial^2 D_q(t)}{\partial t^2} + && \omega_q^2 D_q(t) =
-\frac{2 \omega_q}{\hbar} 
\sum_{\alpha,n,n',{\bf k}}
{\cal{M}}^{\alpha}_{n,n'}(k,q)^{*}
\nonumber \\
&& \times \left\{
N^{\alpha,\alpha}_{n,n'}(k,t) - \delta_{\alpha,v} \ \delta_{n,n'}
\right\} ,
\label{DqEquation}
\end{eqnarray}
subject to the initial conditions
\begin{equation}\
D_q(t=-\infty) = \frac{\partial D_q(t=-\infty)}{\partial t} = 0 . 
\end{equation}

The closed set of coupled partial differential equations,
(\ref{BlochEquations}) and (\ref{DqEquation}), for the carrier density
matrices and coherent phonon amplitudes are converted into a set of coupled
ordinary differential equations (ODE's) by discretizing $k$ and $q$ and
solving for $N^{\alpha,\alpha'}_{n,n'}(k_i)$ and $D(q_i)$ for each of the mesh
points $k_i$ and $q_i$ .
The resulting initial value ODE problem is then solved using a standard
adaptive step-size Runge Kutta routine \cite{NumericalRecipes}.

The phonon distributions do not appear in 
the coupled set of equations (\ref{BlochEquations}) and (\ref{DqEquation}).
If necessary, they can be determined from $N^{\alpha,\alpha}_{n,n'}(k)$
and the pair of equations
\begin{mathletters}
\begin{equation}
\frac{ \partial {\cal{N}}^{coh}_q} {\partial t} =
- \frac{2}{\hbar} \text{Im} \sum_{\alpha,n,n',{\bf k}}
{\cal{M}}^{\alpha}_{n,n'}(k,q) B_q N^{\alpha,\alpha}_{n,n'}(k)
\end{equation}
and
\begin{equation}
\frac{\partial B_q}{\partial t} + i \omega_q B_q =
-\frac{i}{\hbar} \sum_{\alpha,n,n',{\bf k}} {\cal{M}}^{\alpha}_{n,n'}(k,q)^*
N^{\alpha,\alpha}_{n,n'}(k) \ .
\label{BqEquation}
\end{equation}
\end{mathletters}
In Eq.\ (\ref{BqEquation}),\ $B_q(t) \equiv \langle b_q(t) \rangle$
satisfies the initial condition $B_q(t=-\infty)=0$.
For the incoherent phonon distribution,
\begin{equation}
\frac{ \partial {\cal{N}}^{incoh}_q} {\partial t} = 0
\end{equation}
so no incoherent phonons are generated and the incoherent phonon
population maintains its initial thermal equilibrium distribution.

\section{Loaded string model}

The microscopic equations are rather daunting and detailed. In this section,
we show how they can be simplified (under certain conditions) to a more
tractable model, manely that of a driven uniform string, provided one uses
the appropriate driving function, $S(z,t)$, which is nonuniform.
The microscopics, including details of the superlattice band structure
and photogeneration process are included within the driving function.

In our detailed numerical simulations, we use the full microscopic
formalism discussed in the previous sections. However,
we gain a lot of insight if we can deal with the lattice displacement,
$U(z,t)$, directly. If we assume that the acoustic phonon dispersion
relation is linear as in Eq.\ (\ref{PhononDispersion}), then we find that
$U(z,t)$ satisfies the loaded string equation
\begin{equation}
\frac{\partial^2 U(z,t)}{\partial t^2}
- C_s^2 \ \frac{\partial^2 U(z,t)}{\partial z^2} = S(z,t)
\label{WaveEquation}
\end{equation}
subject to the initial conditions
\begin{equation}
U(z,t=-\infty) = \frac{\partial U(z,t=-\infty)}{\partial t} = 0 .
\label{UBoundaryCondition} 
\end{equation}
The LA sound speed, $C_s$, is defined in Eq.\ (\ref{PhononDispersion}),
and the driving function, $S(z,t)$, is given by
\begin{eqnarray}
S(z,t)= && -\frac{1}{\hbar}\sum_{\alpha,n,n'}\sum_{{\bf k},q}
\sqrt{ \frac{2 \hbar \ C_s \ \arrowvert q \arrowvert }{\rho_0 \ V} }
\ {\cal{M}}^{\alpha}_{n',n}(k,q)^{*}
\nonumber \\
&& \times \left\{
N^{\alpha,\alpha}_{n,n'}(k,t) - \delta_{\alpha,v} \ \delta_{n,n'} 
\right\} \ e^{i q z} .
\label{Szt}
\end{eqnarray}

One may question whether a linear phonon dispersion relation is valid in
a superlattice. For small wavevector, $q$, for which
elasticity theory holds, the dispersion relation for LA phonons in 
a superlattice is linear with a dispersion,
$\omega = \overline{C}_s \ q$, where $\overline{C}_s$ is the "average" sound
speed of LA phonons in the well and barriers \cite{yu96}. This, in fact,
has been experimentally verified in
InGaN/GaN superlattice samples studied by C.-K. Sun et. al. \cite{sun00}.

Note that coherent acoustic phonon generation in a superlattice is qualitatively
different than coherent optical phonon generation in a bulk system where
only the $q \approx 0$ optic mode can be excited. As a result, both the
amplitude, $U(z,t)$, and the Fourier transform of the amplitude, $D_q(t)$,
for an optic mode in bulk satisfy a forced oscillator equation. For the
nonuniform, multiple quantum well case, one can excite acoustic modes with
$q \ne 0$. The Fourier transform of the amplitude $D_q(t)$ of a coherent
acoustic phonon obeys a forced oscillator equation, but owing to the linear
dependence of $\omega(q)$ on $q$ ,the amplitude itself, $U(z,t)$, obeys a
1-D wave equation with a forcing term, $S(z,t)$.

Another important point is that  Eq. (\ref{WaveEquation}) can be taken
to be a \textit{uniform} string with a \textit{non-uniform} forcing function. 
This is because the speed of sound is approximately the same in both the $GaN$
and $InGaN$ layers (a more detailed theory would take into account
differences in the sound velocities in each layer).  For propagation of
acoustic modes one can neglect, to lowest order, the differences
between the different layers (this is not true for the optic modes).
The non-uniformity of the forcing function $S(z,t)$ results from differences
in the absorption (not sound velocity) in the well and barrier layers and is
therefore $z$ dependent. We thus see from Eq. (\ref{WaveEquation}) that
understanding coherent acoustic phonons in multiple quantum wells is
equivalent to understanding a
\textit{uniform} string with an \textit{inhomogeneous} forcing term, $S(z,t)$,
containing the microscopics.

To simplify Eq. (\ref{Szt}), we neglect valence band mixing and
assume the effective masses, sound speeds, and coupling constants are
uniform over regions where $S(z,t) \neq 0$, i.e., in regions where carriers
are being photogenerated. We also assume the pump pulses are weak enough so
that screening of the piezoelectric interaction can be neglected. Finally,
if the pump duration is long enough so that transient effects associated with
photogeneration of virtual carriers can be ignored, then the off-diagonal
elements of the carrier density matrices in Eq.\ (\ref{Szt}) can be dropped.
In this case, the driving function takes on the simple form
\begin{equation}
S(z,t) = \sum_{\nu} S_{\nu}(z,t) ,
\label{SztSimple}
\end{equation}
where the summation index, $\nu$, runs over carrier species, i.e.,
conduction electrons, heavy holes, light holes, and crystal field split holes.

Eq.\ (\ref{SztSimple}) says that each carrier species makes a separate
contribution to the driving function.
The partial driving functions, $S_{\nu}(z,t)$, are
\begin{equation}
S_{\nu}(z,t) = \pm \frac{1}{\rho_0} \left\{
a_{\nu} \frac{\partial}{\partial z}+\frac{4 \pi |e| \ e_{33}}{\epsilon_{\infty}}
\right\} \rho_{\nu}(z,t) ,
\label{Snu}
\end{equation}
where the plus sign is used for conduction electrons and the minus sign
is used for holes. Here $\rho_{\nu}(z,t)$ is the
photognerated electron or hole number density, which is real and positive,
and $\rho_0$ is the mass density.
We note that the loaded string equation for the
propagation of coherent phonons together with the simplified driving function
in Eqs.\ (\ref{SztSimple}) and (\ref{Snu})
have also been independently derived by other authors in the limit
$e_{33}=0$. \cite{chigarev00} 

 In Eq.\ (\ref{Snu}), the partial driving function for a given species is
obtained by applying a simple operator to the photogenerated carrier density.
This operator is a sum of two terms, the first due to deformation potential
scattering and the second to piezoelectric scattering. The piezoelectric
coupling constant, $e_{33}$, is the same for all carrier species,
while the deformation potential, $a_{\nu}$, depends on the species. For
conduction electrons, $a_{\nu}=a_{c,z}$, for heavy or light holes,
$a_{\nu}=D_1+D_3$, and for crystal field split holes, $a_{\nu}=D_1$.

It is interesting to note that Planck's constant does not appear
in either the loaded string equation, (\ref{WaveEquation}), or in its 
associated driving function defined in Eqs.\ (\ref{SztSimple})
and (\ref{Snu}). Thus, we find that coherent LA phonon
oscillations in MQW's can be viewed as an essentially classical phenomenon,
an observation that was made in the context of coherent LO phonon
oscillations in bulk semiconductors by Kuznetsov and Stanton in
Ref. \onlinecite{kuznetsov94}.

The driving function, $S(z,t)$, satisfies the sum rule
\begin{equation}
\int_{-\infty}^{\infty} dz \ S(z,t) = 0 .
\label{SumRule}
\end{equation}
This is most easily seen from Eqs.\ (\ref{SztSimple}) and $(\ref{Snu})$,
but it also holds for the general expression in Eq.\ (\ref{Szt}).
The significance of the sum rule is readily appreciated.
After the pump dies away, the carrier density in Eq.\ (\ref{Snu}),
neglecting tunneling between wells, is essentially constant and thus
$S(z,t)$ is time-independent. In the loaded string analogy, the integral
of the driving function over position is proportional to the average force
per unit length on the string. If this integral were non-zero, then the
center of mass of the string would undergo a constant acceleration resulting
in the buildup of an infinite amount of kinetic energy. Such an alarming
result in the context of coherent LA phonons is precluded by the sum rule in
Eq.\ (\ref{SumRule}). 

For a given driving function, the wave equation (\ref{WaveEquation}),
together with the initial conditions (\ref{UBoundaryCondition}), can
be solved for the coherent phonon lattice displacement using the
Green's function method.\cite{berg66} Thus,
\begin{equation}
U(z,t)=\int_{-\infty}^{\infty} dt' \int_{-\infty}^{\infty} dz'
\ G(z-z',t-t') \ S(z',t') .
\label{GreenSolution}
\end{equation}
In our MQW diode model, the substrate is assumed to be infinite and
the Green's function in this case is just
\begin{equation}
G(z,t) = \frac{\Theta(t)}{2 \ C_s} \ \left\{
\Theta(z+C_s \  t) - \Theta(z-C_s \ t) \right\} ,
\end{equation}
where $\Theta(x)$ is the Heaviside step function.

We note that the loaded string model described above is not restricted
to the special case of an infinite substrate and can be extended to study
the generation and propagation of coherent LA phonons in more complicated
heterostructures. If the driving function, $S(z,t)$, due to photoexcited
carriers is localized, then the assumptions leading to Eqs.\ (\ref{SztSimple})
and (\ref{Snu}) need only hold in those regions where $S(z,t)$ in nonvanishing.
The wave equation applies in regions where the LA sound speed, $C_s$, is
constant. Heterostructure in which the LA sound speed is piecewise
constant have abrupt acoustic impedance mismatches which can be handled
by introducing more complicated Green's functions or by using other standard
techniques.\cite{berg66,auld73} An example of such a problem would be
a MQW structure embedded in a free standing substrate in which coherent
LA phonons generated in the MQW could bounce back and forth between two 
parallel substrate-air interfaces.

\section{Results}

In this section, we discuss simulations based on our microscopic theory
of coherent LA phonon generation in a $pin$ diode structure with four periods
of $InGaN/GaN$ MQW's photoexcited by a Gaussian pump normally
incident along the (0001) $z$-direction. The parameters for our numerical
example are listed in Table \ref{SimulationTable}. The MQW dimensions and
Gaussian pump parameters were chosen to match those typically encountered
in room temperature pump-probe differential transmission measurements of
coherent LA phonon oscillations carried out by C.-K. Sun, et. al. \cite{sun00}
on $In_{0.06}Ga_{0.94}N/GaN$ MQW structures having fourteen periods.

\subsection{Bulk wurtzite bandstructure}

\begin{figure}
\begin{center}
\epsfxsize=3.4in
\epsfbox{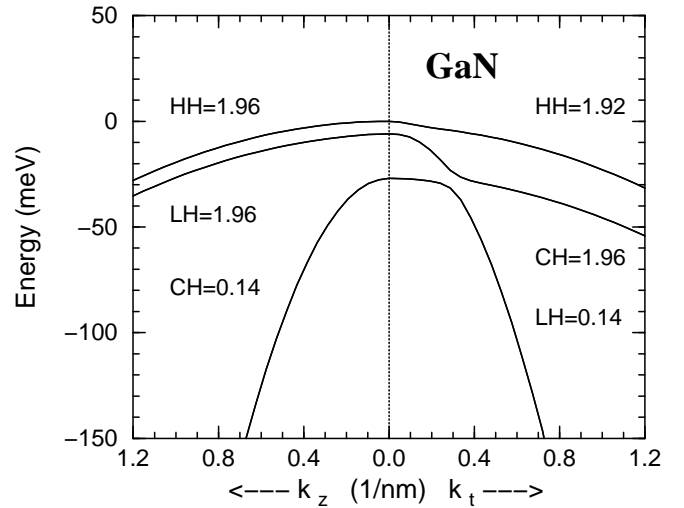}{}
\end{center}
\caption{Bulk $GaN$ valence band structure using effective mass
parameters taken from Table \ref{ParameterTable}. The bands are
plotted along the (0001) $k_z$-axis and along the transverse
$k_t$-axis within the $xy$ plane. The anisotropic zone-center
effective masses for heavy holes (HH), light holes (LH), and crystal
field splitoff holes (CH) are indicated.}
\label{BulkBands}
\end{figure}

Bulk wurtzite $GaN$ and $InN$ are direct gap materials with band
gaps of $3.4$ and 1.95 eV, respectively.
The bulk bandstructure of unstrained wurtzite $GaN$ is shown in
Fig.\ \ref{BulkBands}. As can be seen from equations (\ref{HcMatrix})
and (\ref{HulMatrix}), the bandstructure is anisotropic and depends
on $k_z$, the wavevector along the (0001) $z$-axis, and $k_t$, the
wavevector within the $xy$ plane perpendicular to the $z$-axis. The
effective mass conduction band is two-fold degenerate and has a parabolic
dispersion with anisotropic effective masses $m^{*}_{z}=0.19$ along the
$z$-direction and $m^{*}_{xy}=0.18$ in the $xy$ plane.

The two-fold degenerate valence bands are mixtures of heavy hole (HH),
light hole (LH), and crystal-field splitoff hole (CH) character. At the
zone center, the off-diagonal components of the $3 \times 3$ upper and
lower Hamiltonians in Eq.\ (\ref{HulMatrix}) vanish and the valence bands
can be labeled according to their pure state wavefunctions at $k=0$.
For the zone-center HH state, the degenerate wavefunctions are the
basis states $\arrowvert v,1 \rangle$ and $\arrowvert v,4 \rangle$
defined in Eq.\ (\ref{vBasis}). For the zone-center LH state, the
wavefunctions are $\arrowvert v,2 \rangle$ and $\arrowvert v,5 \rangle$,
and for the CH band the zone-center wavefunctions are
$\arrowvert v,3 \rangle$ and $\arrowvert v,6 \rangle$.
The heavy-hole effective masses along $z$ and $xy$ are
$m^{HH}_{z}=\arrowvert A_1+A_3 \arrowvert^{-1}=1.96$
and $m^{HH}_{xy}=\arrowvert A_2+A_4-A_5 \arrowvert^{-1}=1.92$
for heavy holes, $m^{LH}_{z}=m^{HH}_{z}=1.96$
and $m^{LH}_{xy}=\arrowvert A_2+A_4+A_5 \arrowvert^{-1}=0.14$
for light holes, and $m^{CH}_{z}=\arrowvert A_1 \arrowvert^{-1}=0.14$
and $m^{CH}_{xy}=\arrowvert A_2 \arrowvert^{-1}=1.96$ for crystal-field
splitoff holes.

\subsection{Pseudomorphic strain}

\begin{figure}
\begin{center}
\epsfxsize=3.4in
\epsfbox{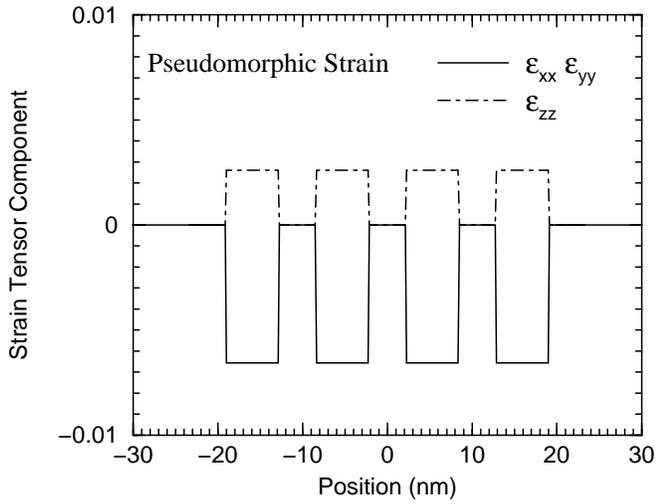}{}
\end{center}
\caption{Strain tensor components for pseudomorphically strained
$In_xGa_{1-x}N$ multi-quantum well diode as a function of position. The 
diode parameters are listed in Table \ref{SimulationTable}. }
\label{StrainFigure}
\end{figure}

Bulk $GaN$ and $InN$ have different lattice constants so when an (0001)
$InGaN$ MQW structure is grown, a significant
lattice mismatch occurs between the $In_xGa_{1-x}N$ wells and $GaN$ barriers. 
For the $InGaN$ MQW diode specified in Table \ref{SimulationTable}, we
assume pseudomorphic strain conditions. In a pseudomorphically strained
device, the lattice constant throughout the MQW adjusts to the value of
the lattice constant in the bulk $N$ and $P$ substrates in order to
minimize the overall strain energy. In our simulated diode, the substrates
are $n$- and $p$- doped $GaN$, so the lattice constant throughout the device
takes on the $GaN$ value, i.e. $a_0=3.189 \ \AA$.
The non-vanishing position-dependent strain tensor components,
$\epsilon_{xx}$, $\epsilon_{yy}$, and $\epsilon_{zz}$, for the MQW diode,
as computed from Eqs.\ (\ref{Epsilonxx}) and (\ref{Epsilonzz}),
are shown in Fig.\ \ref{StrainFigure} as a function of $z$. Clearly, the
$GaN$ barriers are unstrained since the $N$ and $P$ substrates are
composed of $GaN$ and all the strain from the lattice mismatch is
accomodated in the $In_{0.06}Ga_{0.94}N$ wells.

\subsection{Built-in piezoelectric field}

\begin{figure}
\begin{center}
\epsfxsize=3.4in
\epsfbox{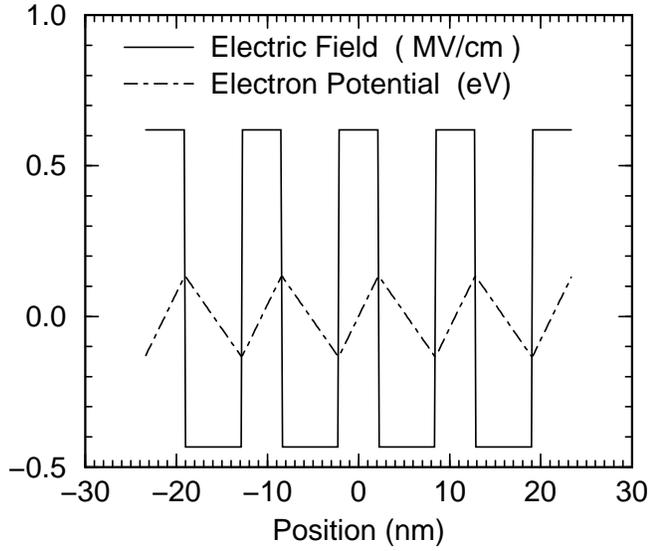}{}
\end{center}
\caption{Electric field and potential for the strain field in 
Fig. \ref{StrainFigure}. The applied dc bias, $V_A$, has been adjusted so
flat-band biasing is achieved, i.e. so that the band edges are periodic
functions of position. The diode parameters are listed in
Table \ref{SimulationTable}. }
\label{Piezofield}
\end{figure}

The presence of strain in the MQW's results in the creation of
a strain-induced polarization, $P^0_z(z)$, directed along $z$ as described
by Eq.\ (\ref{PzEq}). The strain-induced polarization, in turn, results in   
a strong bult-in piezoelectric field which can be computed from
Eqs.\ (\ref{EzEq}) and (\ref{P0Eq}), given the strain field and the 
dc bias, $V_A$, applied across the diode.
The computed strain-induced piezoelectric field, $E^0_z(z)$,
and the piezoelectric confinement potential, $V_{\text{piezo}}(z)$,
which result from the strain field in Fig. \ref{StrainFigure} are shown in
Fig. \ref{Piezofield}. Prior to the application of the pump pulse, we assume
the applied dc bias, $V_A$, has been adjusted so flat-band biasing in
the diode is achieved, i.e. $V_A$ is such that the band edges seen in
Fig.\ \ref{Piezofield} are periodic functions of position.

\begin{figure}
\begin{center}
\epsfxsize=3.4in
\epsfbox{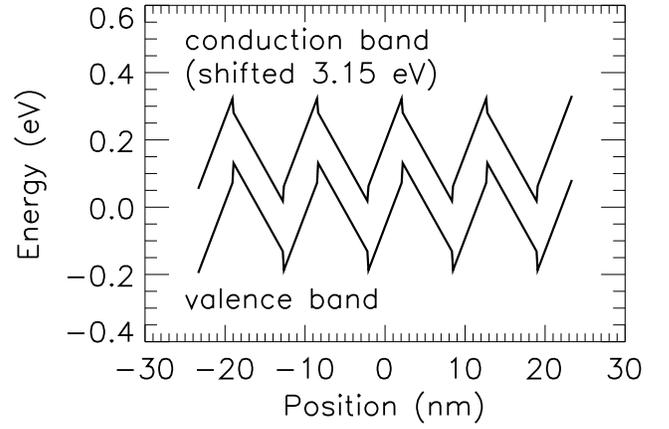}{}
\end{center}
\caption{Conduction and valence band edges for pseudomorphically strained
$In_xGa_{1-x}N$ multi-quantum well diode as a function of position.
The applied dc bias, $V_A$, has been adjusted so flat-band biasing
is achieved, i.e. so that the band edges are periodic functions of position.
The diode parameters are listed in Table \ref{SimulationTable}. }
\label{BandedgeFigure}
\end{figure}

Given the piezoelectric field and confinement potentials, position
dependent band edges for the MQW can be computed.
The conduction and valence band edges for our pseudomorphically strained
MQW diode are shown as functions of position in Fig.\ 
\ref{BandedgeFigure}. These are just the confinement potentials,
$V_{\alpha}(z)=V_{\alpha,\text{gap}}(z)+V_{\text{piezo}}(z)$, in the
diode prior to photoexcitation.
It is clear from Fig.\ \ref{BandedgeFigure} that the confinement of elecrons
and holes in the MQW is mostly due to strong built-in piezoelectric
fields which result in the triangular confinement potentials seen in each
well.

\subsection{Photogeneration of carriers}

In our numerical example, we simulate photoexcitation of electrons
and holes and the generation and subsequent propagation of coherent LA
phonons in the hypothetical MQW diode when a Gaussian pump laser pulse
is normally incicent along the $z$-axis.
As seen in Table \ref{SimulationTable}, the Gaussian pump pulse is assumed
to be left circularly polarized with a photon energy of $3.21\ \text{eV}$.
The pump fluence is taken to be $100.0\ \mu \text{J}/\text{cm}^2$
and the Gaussian FWHM is taken to be $180.0\ \text{fs}$. The expermient
is assumed to take place at room temperature.

\begin{figure}
\begin{center}
\epsfxsize=3.4in
\epsfbox{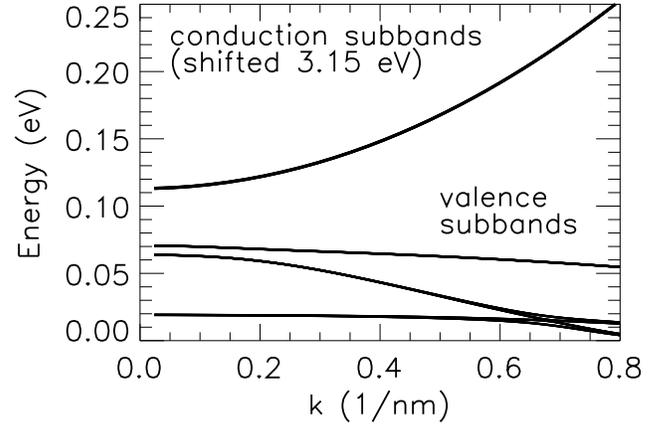}{}
\end{center}
\caption{Conduction and valence subband energies as functions
of $k$ for the $In_xGa_{1-x}N$ diode structure described in
Table \ref{SimulationTable}.}
\label{SubbandFigure}
\end{figure}

In Fig.\ \ref{SubbandFigure}, the computed conduction and valence subband
energies are shown as functions of $k$ for the $In_xGa_{1-x}N$ diode. 
At the chosen pump energy of $3.21\ eV$, electrons from the first
two valence subbands are excited into the lowest lying conduction subband.

\begin{figure}
\begin{center}
\epsfxsize=3.4in
\epsfbox{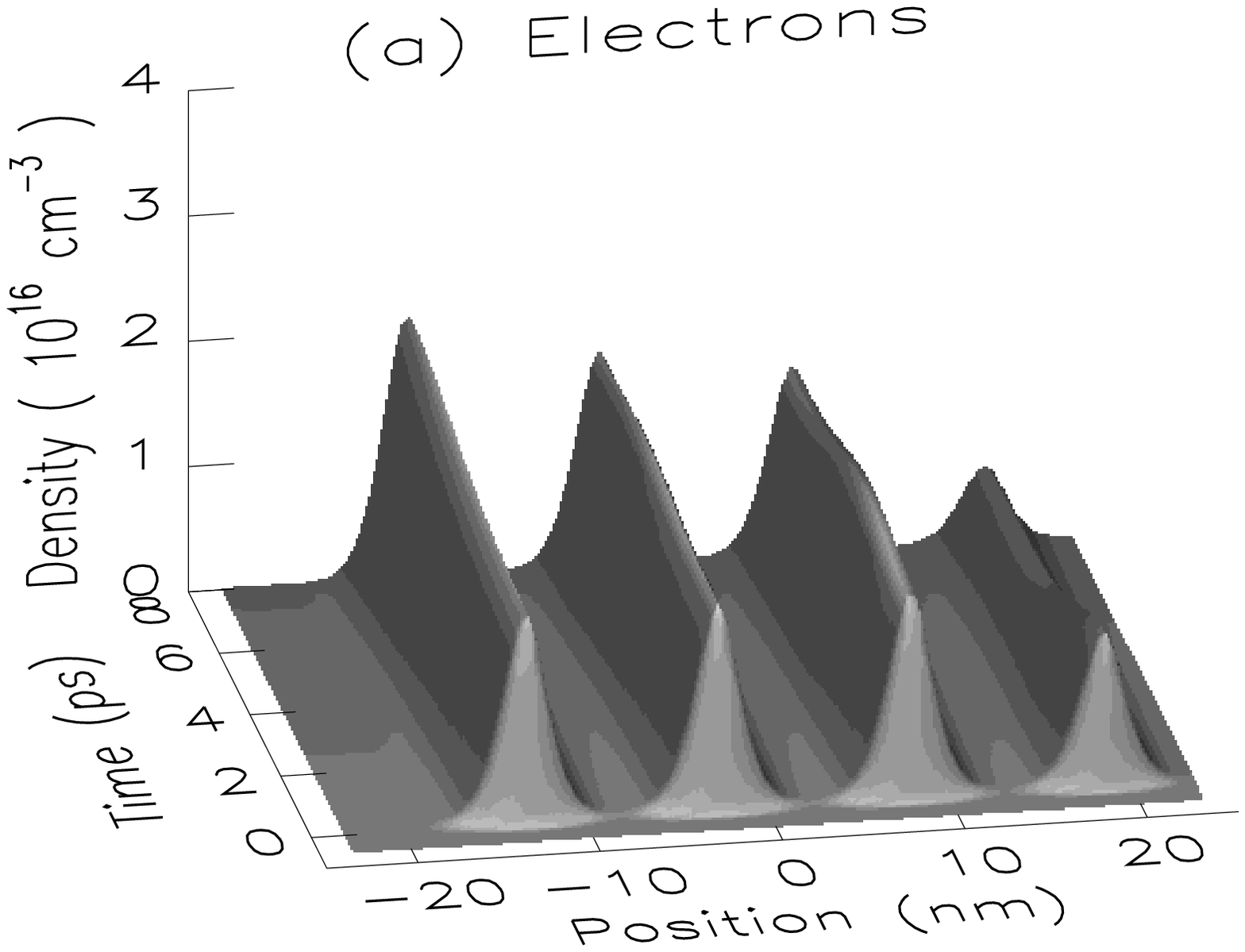}{}
\epsfxsize=3.4in
\epsfbox{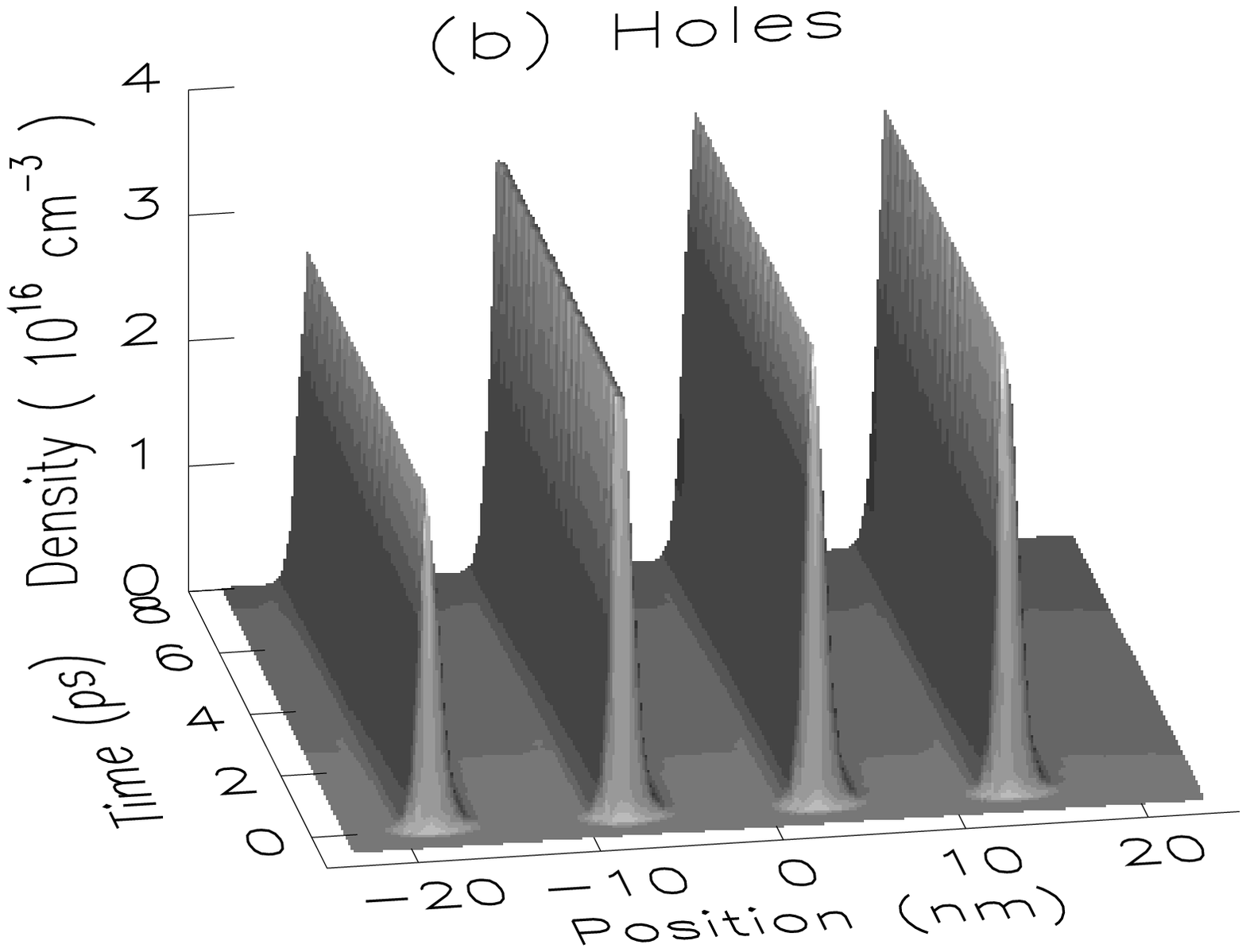}{}
\end{center}
\caption{Density of excited carriers computed in the absence
of Coulomb effects for (a) electrons and (b) holes as functions of position
for the $In_xGa_{1-x}N$ diode structure and laser pumping parameters shown
in Table \ref{SimulationTable}. }
\label{3dDensity1}
\end{figure}

\begin{figure}
\begin{center}
\epsfxsize=3.4in
\epsfbox{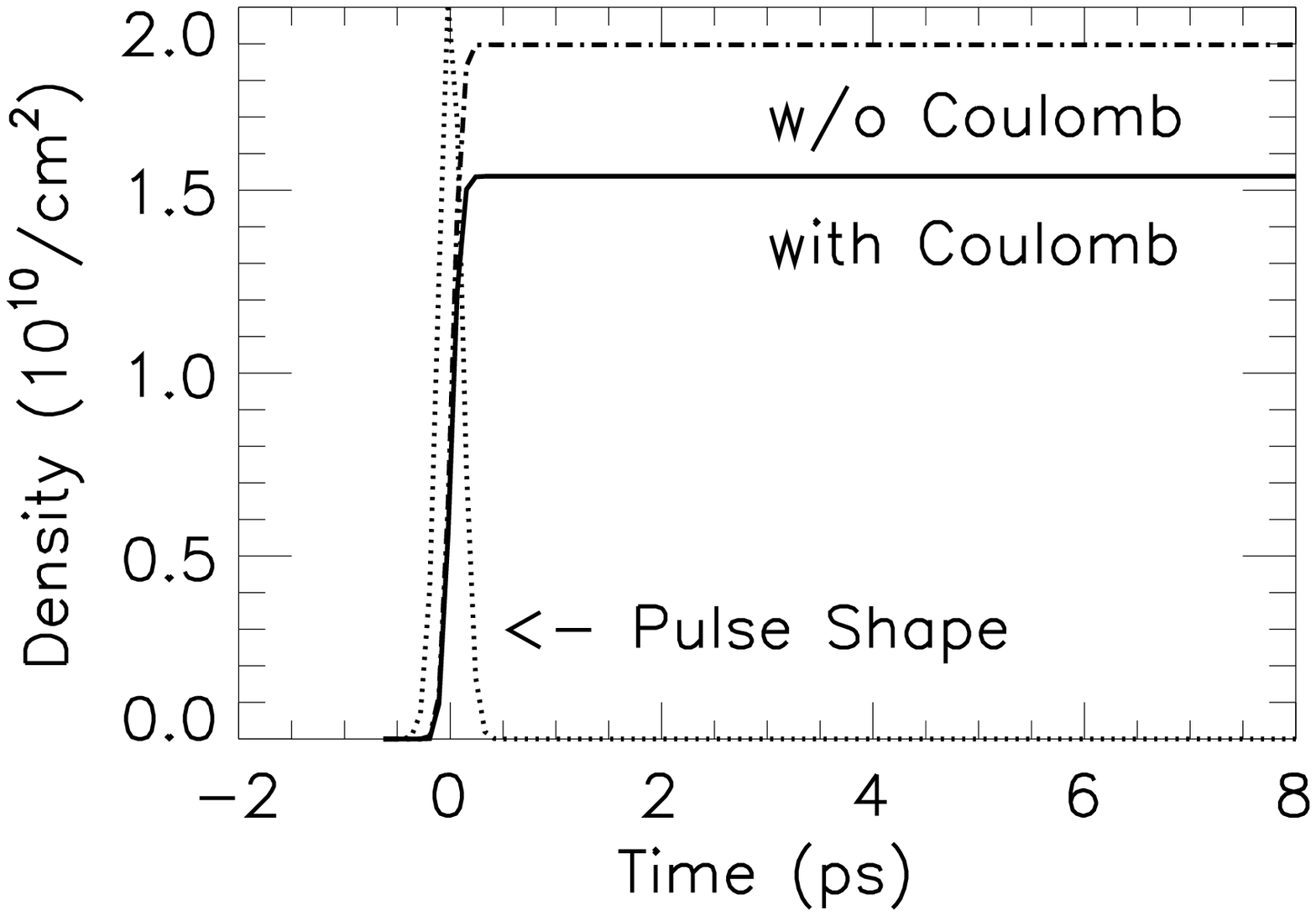}{}
\end{center}
\caption{Total photoexcited electron density with and without Coulomb
effects as a function of time for the $In_xGa_{1-x}N$ diode structure and
laser pumping parameters listed in Table \ref{SimulationTable}. The pulse
shape (arb. units) is shown for comparison. }
\label{TotalDensity}
\end{figure}

The computed densities of photoexcited electrons and holes, neglecting
and including Coulomb interaction effects, are shown as functions of
position and time in Figs.\ \ref{3dDensity1} and \ref{3dDensity2},
respectively and the total photoexcited electron density per unit
area as a function of time is shown in Fig.\ \ref{TotalDensity}.
In Fig.\ \ref{TotalDensity} the pulse shape is shown for comparison. We find 
that including Coulomb effects decreases the total photogenerated carrier
density. The electrons and holes screen the built-in piezoelectric field
widening the effective band gap. This quantum confined Stark effect acts to
suppress the photogeneration of carriers.

\subsection{Generation of coherent phonons}

\begin{figure}
\begin{center}
\epsfxsize=3.4in
\epsfbox{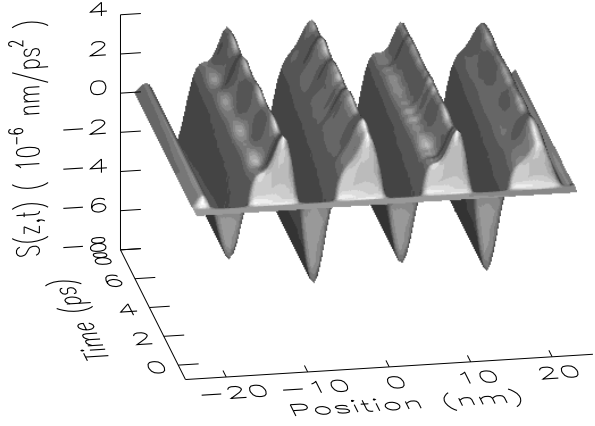}{}
\end{center}
\caption{Driving function, $S(z,t)$, for the coherent LA phonon wave
equation as a function of position and time for the $In_xGa_{1-x}N$
diode structure and laser pumping parameters in Table \ref{SimulationTable}.
$S(z,t)$ is computed using the full microscopic expression of
Eq.\ (\ref{Szt}).}
\label{DrivingFunction}
\end{figure}

The driving function, $S(z,t)$, for the driven string equation
(\ref{WaveEquation}) is shown in Fig.\ \ref{DrivingFunction}
as a function of position and time. The driving function has units
of acceleration and in Fig.\ \ref{DrivingFunction}, we compute $S(z,t)$
using the full microscopic formalism of Eq.\ (\ref{Szt}).

\begin{figure}
\begin{center}
\epsfxsize=3.4in
\epsfbox{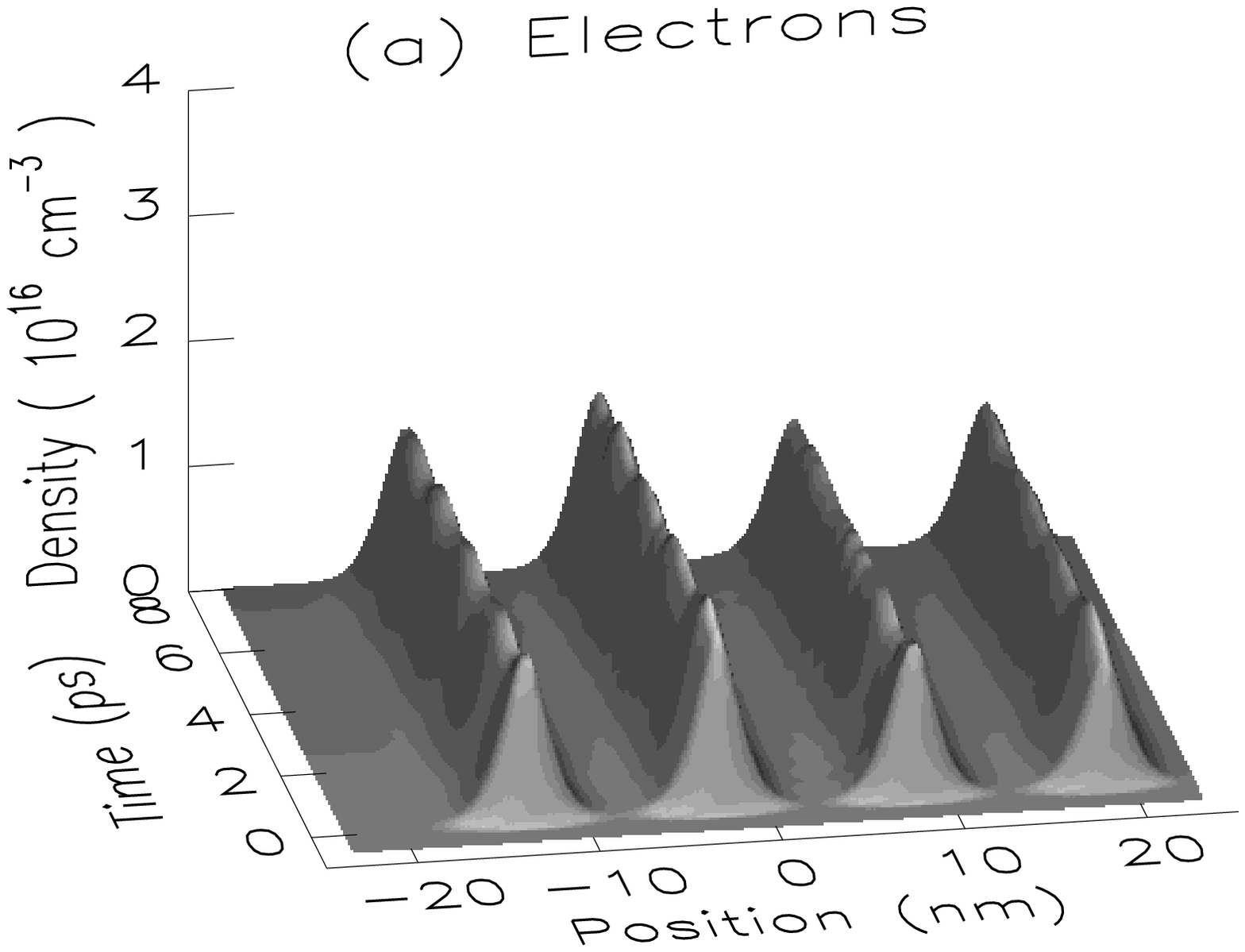}{}
\epsfxsize=3.4in
\epsfbox{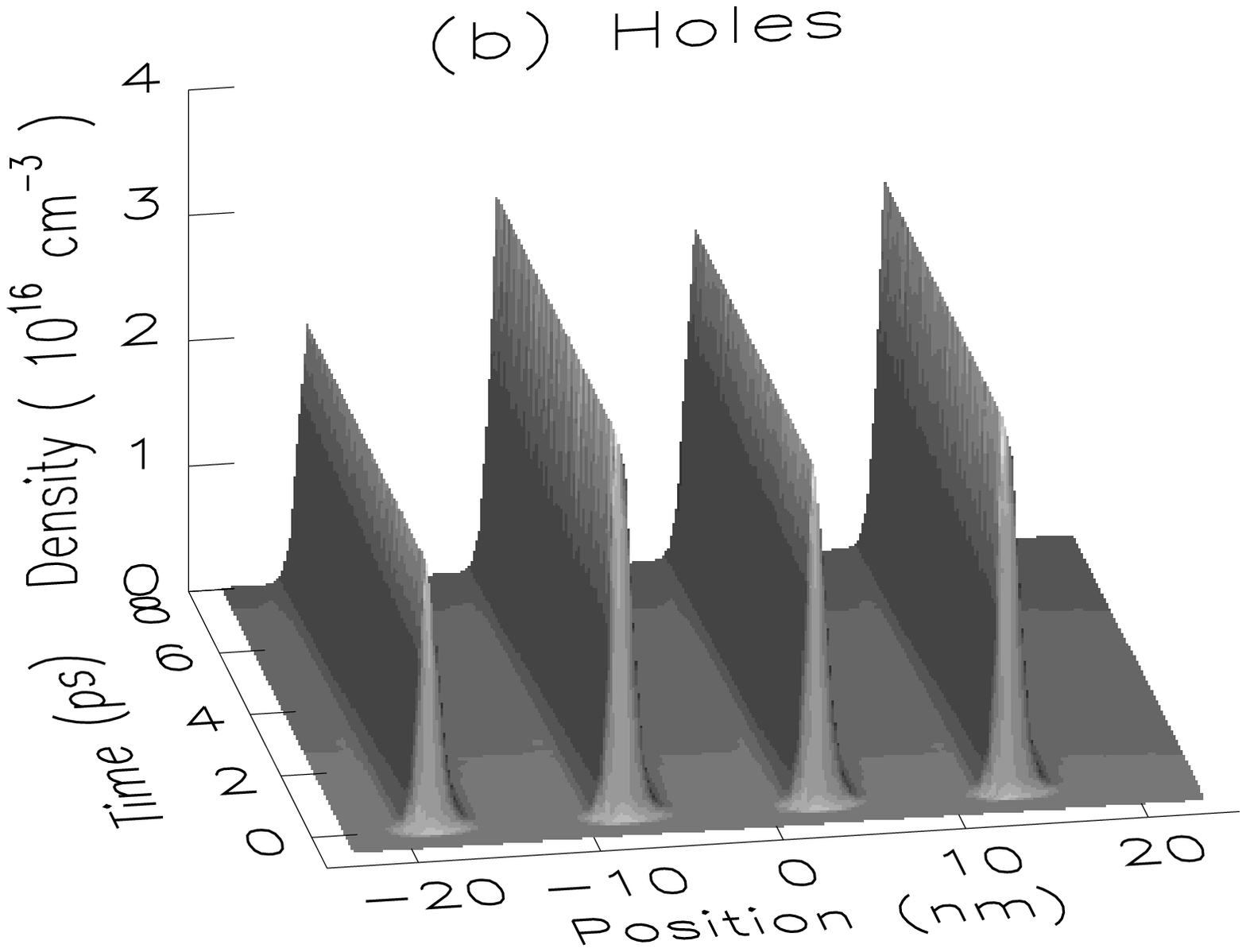}{}
\end{center}
\caption{Density of excited carriers including
 Coulomb effects for (a) electrons and (b) holes as functions of position
for the $In_xGa_{1-x}N$ diode structure and laser pumping parameters taken
from Table \ref{SimulationTable}. }
\label{3dDensity2}
\end{figure}

\begin{figure}
\begin{center}
\epsfxsize=3.4in
\epsfbox{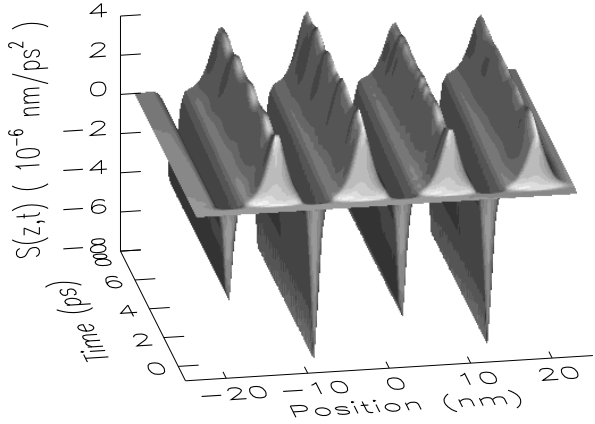}{}
\end{center}
\caption{Driving function, $S(z,t)$, in the simplified 
loaded string model for the coherent LA phonon wave
equation as a function of position and time for the $In_xGa_{1-x}N$
diode structure and laser pumping parameters in Table \ref{SimulationTable}.}
\label{DrivingFunction2}
\end{figure}

For comparison, we also computed the driving function in the simplified
loaded string model of Eqs.\ (\ref{SztSimple}) and (\ref{Snu}) using
the carrier densities shown in Fig.\ \ref{3dDensity2} to facilitate
the comparison. Since the photoexcited
holes are predominantly a mixture of heavy- and light- holes, we use 
$a_{\nu}=D_1+D_3$ in computing hole deformation potential contributions in
Eq.\ (\ref{Snu}). The sum over species, $\nu$, then yields the total 
driving function
\begin{eqnarray}
S(z,t) && = \frac{1}{\rho_0} \left\{
a_{c,z} \frac{\partial}{\partial z}+\frac{4 \pi |e| e_{33}}{\epsilon_{\infty}}
\right\} \rho_{\text{elec}}(z,t)
\nonumber \\
&& - \ \frac{1}{\rho_0} \left\{
(D_1+D_3) \frac{\partial}{\partial z}+\frac{4 \pi |e| e_{33}}{\epsilon_{\infty}}
\right\} \rho_{\text{hole}}(z,t) ,
\label{Snu2}
\end{eqnarray}
where $\rho_{\text{elec}}(z,t)$ and $\rho_{\text{hole}}(z,t)$ are the 
total conduction electron and valence hole densities plotted in
Fig.\ \ref{3dDensity2}. The resulting $S(z,t)$ is shown in
Fig.\ \ref{DrivingFunction2}.

By comparing Figs.\ \ref{DrivingFunction}
and \ref{DrivingFunction2}, we see that for the diode structure
and Gaussian pump used in our simulation the simplified loaded
string model produces essentially the same results as those
obtained using the full microscopic formalism. 

Acoustic LA phonon generation due to the piezoelectric 
effect depends on the piezoelectric constant, $e_{33}$, the
number of photogenerated electrons and holes, as
well as the spatial separation of electron and hole densities brought
about by the strong built-in piezoelectric field in the MQW's. From
Eq.\ (\ref{Snu2}), the piezoelectric contribution to the driving 
function is given by
\begin{equation}
S_{\text{piezo}}(z,t) = \frac{1}{\rho_0}\ 
\frac{4 \pi |e| \ e_{33}}{\epsilon_{\infty}}
\left\{ \rho_{\text{elec}}(z,t)- \rho_{\text{hole}}(z,t) \right\}
\label{Spiezo}
\end{equation}
In the absence of a built-in piezoelectric field (such as found
in a square well with infinite barriers) we would have
$\rho_{\text{elec}}(z,t) \approx \rho_{\text{hole}}(z,t)$
and hence $S_{\text{piezo}}(z,t) \approx 0$ even for relatively large
values of $e_{33}$. The built-in piezoelectric field serves to 
spatially separate the electrons and holes so that
$\rho_{\text{elec}}(z,t) \neq \rho_{\text{hole}}(z,t)$
and hence $S_{\text{piezo}}(z,t) \neq 0$. However,
if the built-in piezoelectric field is too strong and the
spatial separation of electrons and holes too large, then
$\rho_{\text{elec}}(z,t) - \rho_{\text{hole}}(z,t) \approx 0$.
This is because the overlap between the conduction and valence envelope
functions enters into the optical dipole matrix elements in Eq.\ (\ref{dcv}).
If there's negligable overlap between electron and hole envelope functions
due to strong piezoelectric fields then
${\bf{d}}^{cv}_{n,n'}({\bf{k}}) \approx 0$,
no electron-hole pairs are photogenerated and once again
$S_{\text{piezo}}(z,t) \approx 0$.

The deformation potential contribution to the driving function is given by
\begin{equation}
S_{\text{def}}(z,t) =
\frac{a_{c,z}}{\rho_0} \frac{\partial \rho_{\text{elec}}(z,t)}{\partial z}
-\frac{(D_1+D_3)}{\rho_0} \frac{\partial \rho_{\text{hole}}(z,t)}{\partial z}. 
\label{Sdef}
\end{equation}
From Table \ref{ParameterTable}, the conduction electron deformation
potential, $a_{c,z}$, is roughly twice the valence hole
deformation potential, $D_1+D_3$. Thus, the two terms in Eq.\ (\ref{Sdef})
are of comparable magnitude. The first term, due to conduction electrons,
gives rise to a contribution to $S_{\text{def}}(z,t)$ which is 
localized on the right side of each MQW while the second, due to 
valence holes, gives rise to contribution which is localized on the left
hand side of each MQW.

\begin{figure}
\begin{center}
\epsfxsize=3.4in
\epsfbox{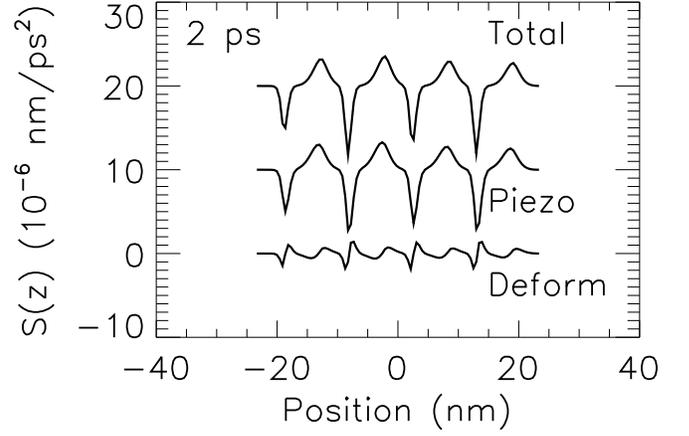}{}
\end{center}
\caption{Driving function, $S(z,t)$, in the simplified loaded string
model at $t=2\ \text{ps}$ for the coherent LA phonon wave equation
as a function of position for the $In_xGa_{1-x}N$ diode structure
and laser pumping parameters in Table \ref{SimulationTable}. The
total driving function, $S(z,t)$, is the sum of piezoelectric and
deformation potential contributions, $S_{\text{piezo}}(z,t)$
and $S_{\text{def}}(z,t)$. }
\label{SztString}
\end{figure}

In our simulation, we find that piezoelectric and deformation potential
contriburtions to the driving function are comparable. This
is seen in Fig. \ref{SztString} where $S_{\text{piezo}}(z)$ and
$S_{\text{def}}(z)$, along with their sum, are plotted at $t=2\ \text{ps}$.
In this example, we find that $S_{\text{piezo}}(z)$ makes the dominant
contribution to $S(z,t)$ as can be seen in Fig.\ \ref{SztString}. 

\begin{figure}
\begin{center}
\epsfxsize=3.4in
\epsfbox{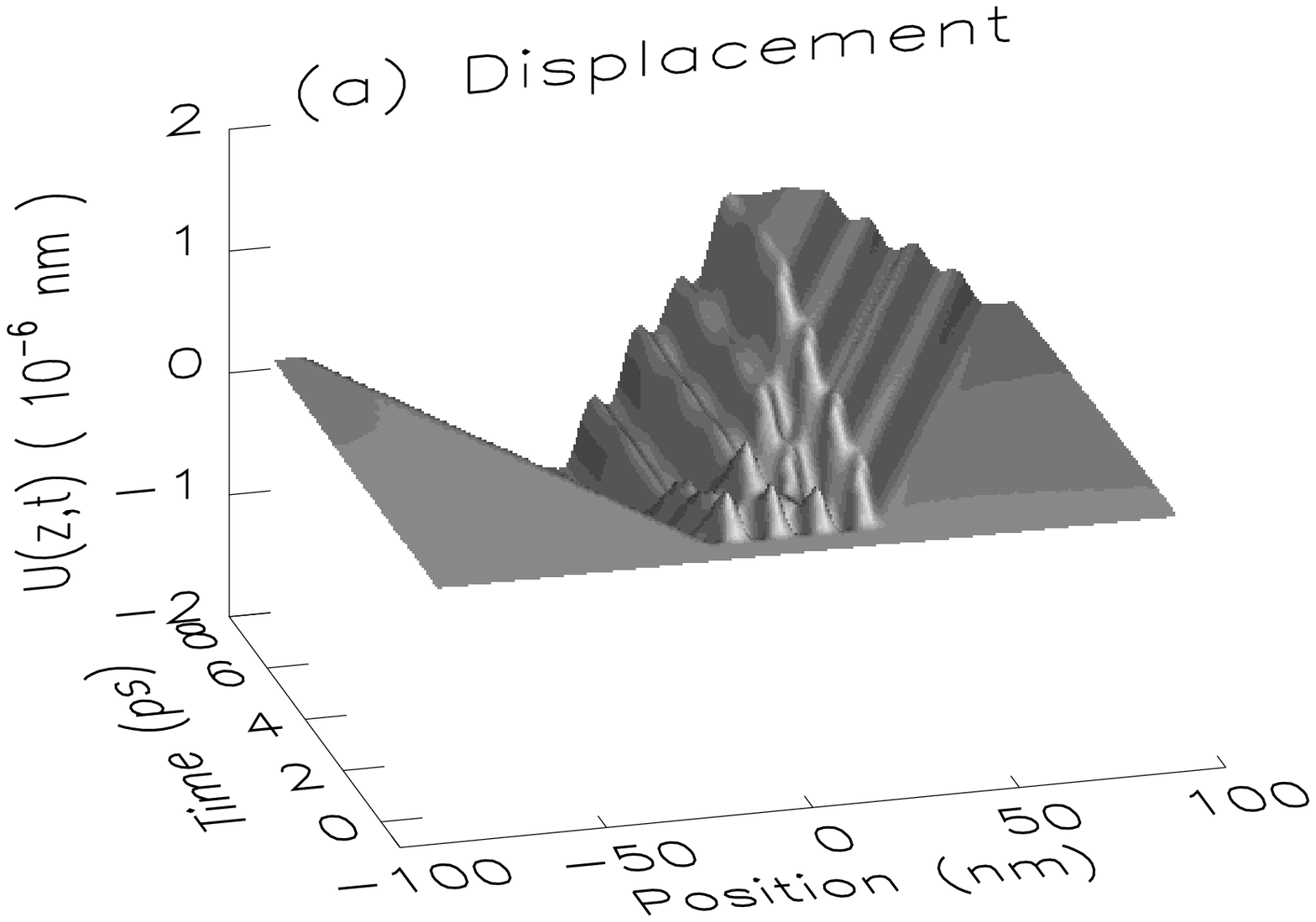}{}
\epsfxsize=3.4in
\epsfbox{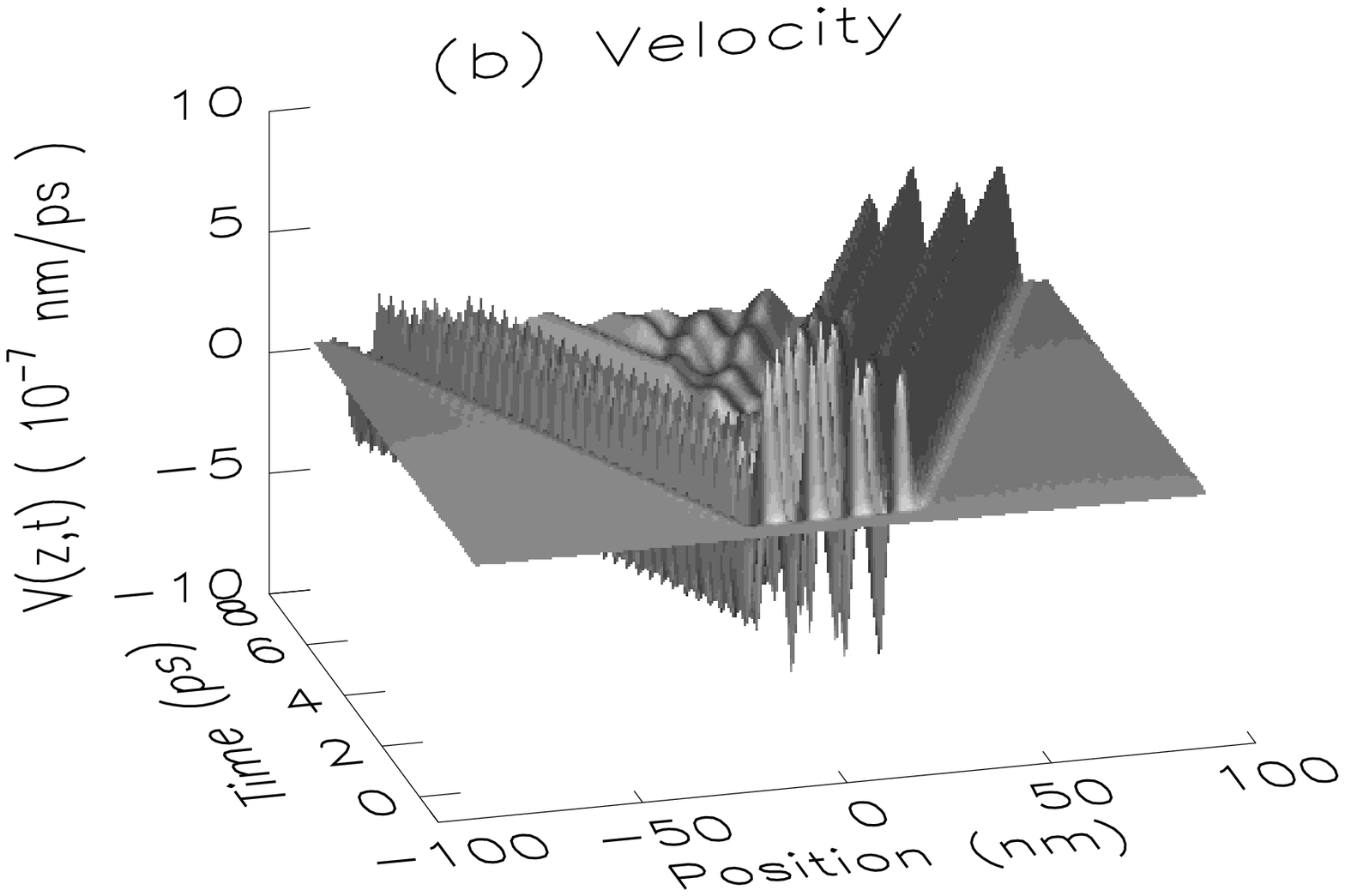}{}
\end{center}
\caption{Lattice displacement,
$U(z,t)$, and velocity field, $V(z,t)$, for coherent LA phonons
generated by the driving function shown in Fig. \ref{DrivingFunction}. }
\label{UVFigure}
\end{figure}

The macroscopic lattice displacement, $U(z,t)$, and velocity field,
$V(z,t)$, can be obtained from the coherent phonon amplitudes, $D_q(t)$,
using Eqs.\ (\ref{Uzt}) and (\ref{Vzt}).
Alternatively, they can also be obtained from the driving
function, $S(z,t)$, through the Green's function solution of the driven
string equation.
In Fig. \ref{UVFigure}, we plot the macroscopic lattice displacement,
$U(z,t)$, and velocity field, $V(z,t)$, for coherent LA phonon modes
generated by the driving function shown in Fig. \ref{DrivingFunction}.

\subsection{Coherent phonon energy}

From the lattice displacement, $U(z,t)$, we can obtain the total energy
density per unit volume associated with coherent LA phonons,
\begin{mathletters}
\label{EnergyDensityEquations}
\begin{equation}
{\cal{E}}_{LA}(z,t)={\cal{T}}_{LA}(z,t)+{\cal{V}}_{LA}(z,t) ,
\label{TotalEnergyDensity}
\end{equation}
as the sum of a kinetic energy density term,
\begin{equation}
{\cal{T}}_{LA}(z,t)= \frac{\rho_0(z)}{2}
\left( \frac{\partial U(z,t)}{\partial t} \right)^2 ,
\label{KineticEnergyDensity}
\end{equation}
and a potential energy density term,
\begin{equation}
{\cal{V}}_{LA}(z,t)= \frac{C_{33}(z)}{2} 
\left( \frac{\partial U(z,t)}{\partial z} \right)^2 .
\label{PotentialEnergyDensity}
\end{equation}
\end{mathletters}
The coherent LA phonon energy per unit area, $E_{LA}(t)$,  
is obtained by integrating ${\cal{E}}_{LA}(z,t)$ over position, $z$:
\begin{equation}
E_{LA}(t)=\int_{-\infty}^{\infty} dz \ {\cal{E}}_{LA}(z,t) .
\label{ELAz}
\end{equation}

\begin{figure}
\begin{center}
\epsfxsize=3.4in
\epsfbox{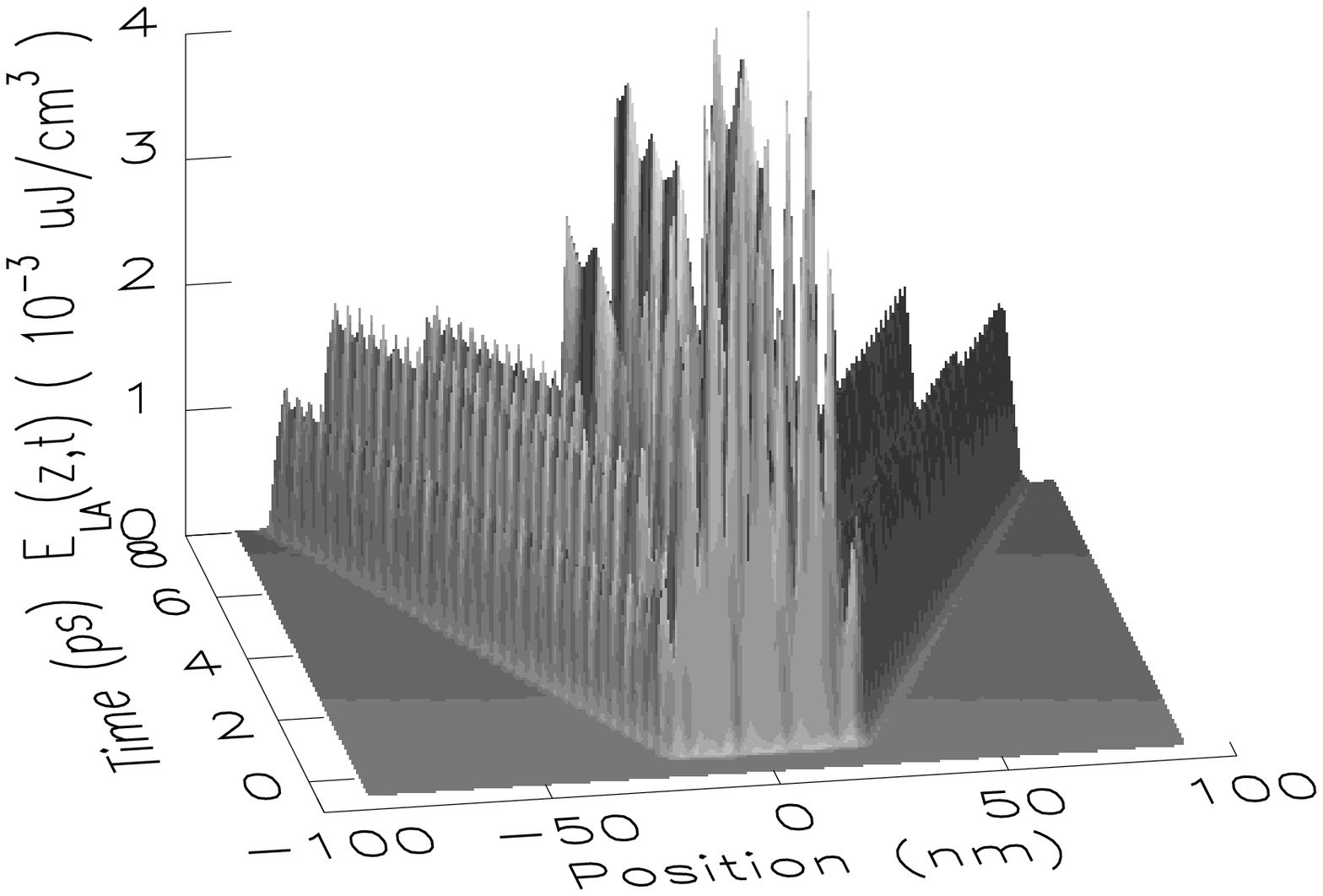}{}
\epsfxsize=3.4in
\epsfbox{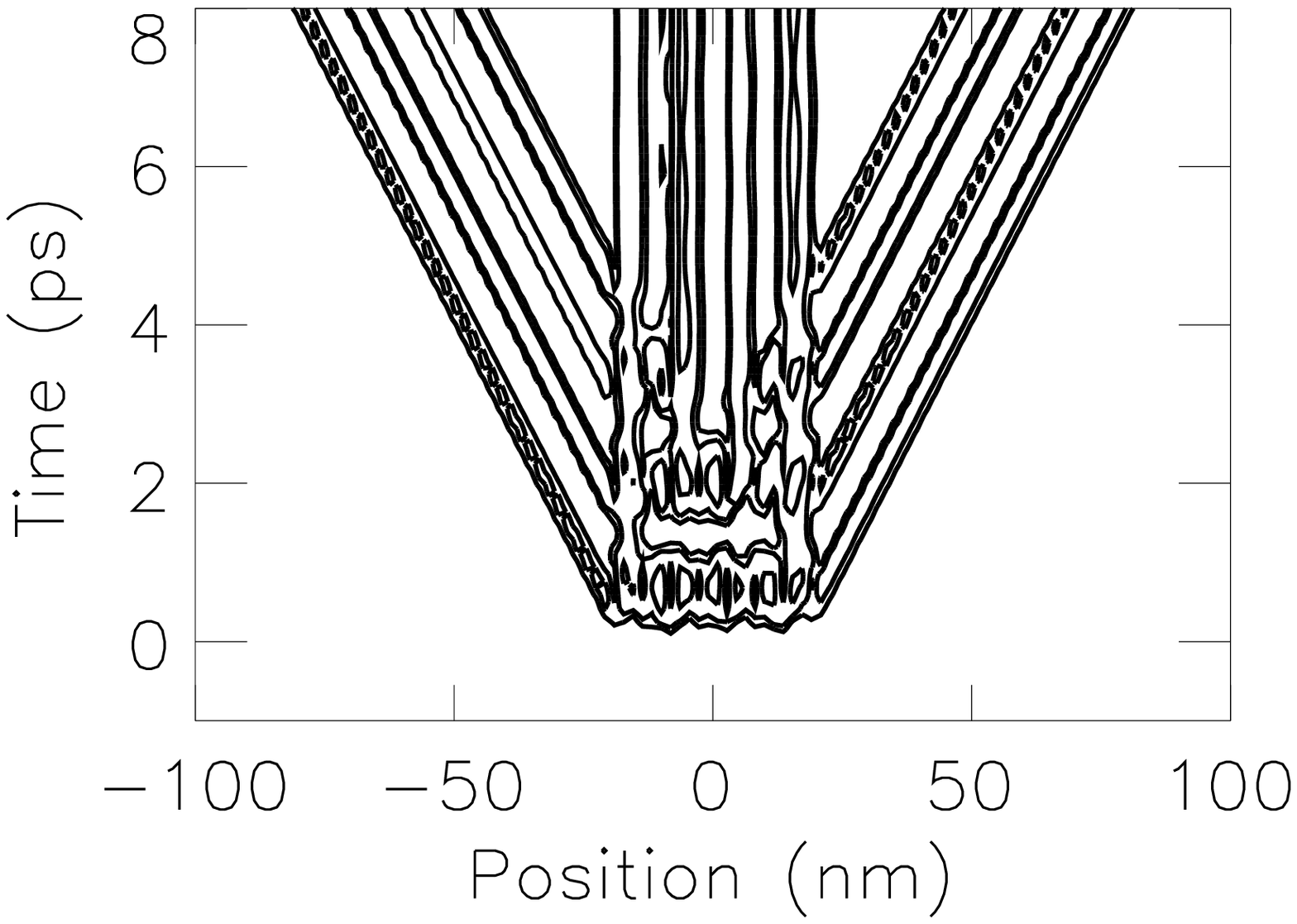}{}
\end{center}
\caption{Total energy density, ${\cal{E}}_{LA}(z,t)$,
for coherent LA phonons as a function of
position and time for the driving function shown in
Fig. \ref{DrivingFunction}. The total integrated energy density as 
a function of time is obtained by integrating over position, $z$. }
\label{EnergyDensity}
\end{figure}

\begin{figure}
\begin{center}
\epsfxsize=3.4in
\epsfbox{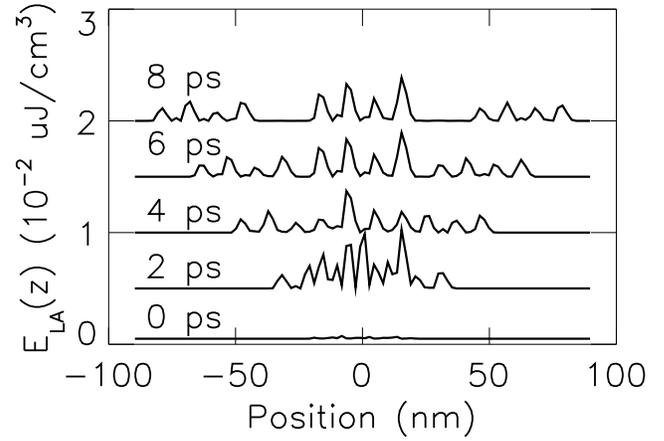}{}
\end{center}
\caption{Total energy density, ${\cal{E}}_{LA}(z,t)$,
for coherent LA phonons as a function of
position for several values of the time, $t$, for the driving function shown
in Fig. \ref{DrivingFunction}. The curves for different times have been
offset to avoid overlapping.}
\label{EnergyDensityMovie}
\end{figure}

The total energy density, ${\cal{E}}_{LA}(z,t)$, for coherent LA phonons as a
function of position and time is shown in Fig. \ref{EnergyDensity} and in
Fig. \ref{EnergyDensityMovie} the total energy density is plotted as
a function of position for equally spaced values of the time ranging from
$t=0$ to $t=8\ \text{ps}$ in increments of $2\ \text{ps}$.
The curves in Fig. \ref{EnergyDensityMovie} for different times have been
offset to avoid overlapping. At short times, the evolution of the total phonon
energy density is complicated, but the long times behavior,
$t\ \gtrsim \ 6\ \text{ps}$, can be easily understood.

As $t \rightarrow \infty$, a localized energy density appears in the MQW
region due almost entirely to the potential energy term in
Eq.\ (\ref{PotentialEnergyDensity}). This is due to near steady-state loading
by the driving function at long times.
Assuming the driving function, $S(z,t)$, is approximately constant at long
times, the loaded string equation, (\ref{WaveEquation}), can be integrated
once in the steady state limit. We find the steady state solution
\begin{equation}
\frac{\partial U(z)}{\partial z} \approx
- \int_{-\infty}^{z} dz' \ \frac{S(z')}{C_s^2} .
\label{StaticSolution} 
\end{equation}
from which the long time behavior of the LA phonon energy density per unit
volume in the MQW's,
\begin{equation}
{\cal{E}}_{LA}^{\infty}(z) \approx \frac{C_{33}(z)}{2\ A} 
\left(
\int_{-\infty}^{z} dz' \ \frac{S(z')}{C_s^2}
\right)^2 ,
\label{StaticEnergyDensity}
\end{equation}
can be obtained. The fact that the energy density in
Eq.\ (\ref{StaticEnergyDensity}) is localized in the MQW's follows
directly from the sum rule (\ref{SumRule}) and is clearly seen in
Figs. \ref{EnergyDensity} and \ref{EnergyDensityMovie}.

In addition to the localized energy density, which remains behind in the
MQW's, two propagating wave trains consisting of four pulses each are seen to
exit the MQW region and travel off to infinity at the acoustic phonon sound
speed, $C_s$. The distance between the pulses is just the inter-well
separation distance. In these radiating wave trains, the kinetic and potential
energy densities, ${\cal{T}}_{LA}(z,t)$ and ${\cal{V}}_{LA}(z,t)$,
are found to be equal as one would expect.

\begin{figure}
\begin{center}
\epsfxsize=3.4in
\epsfbox{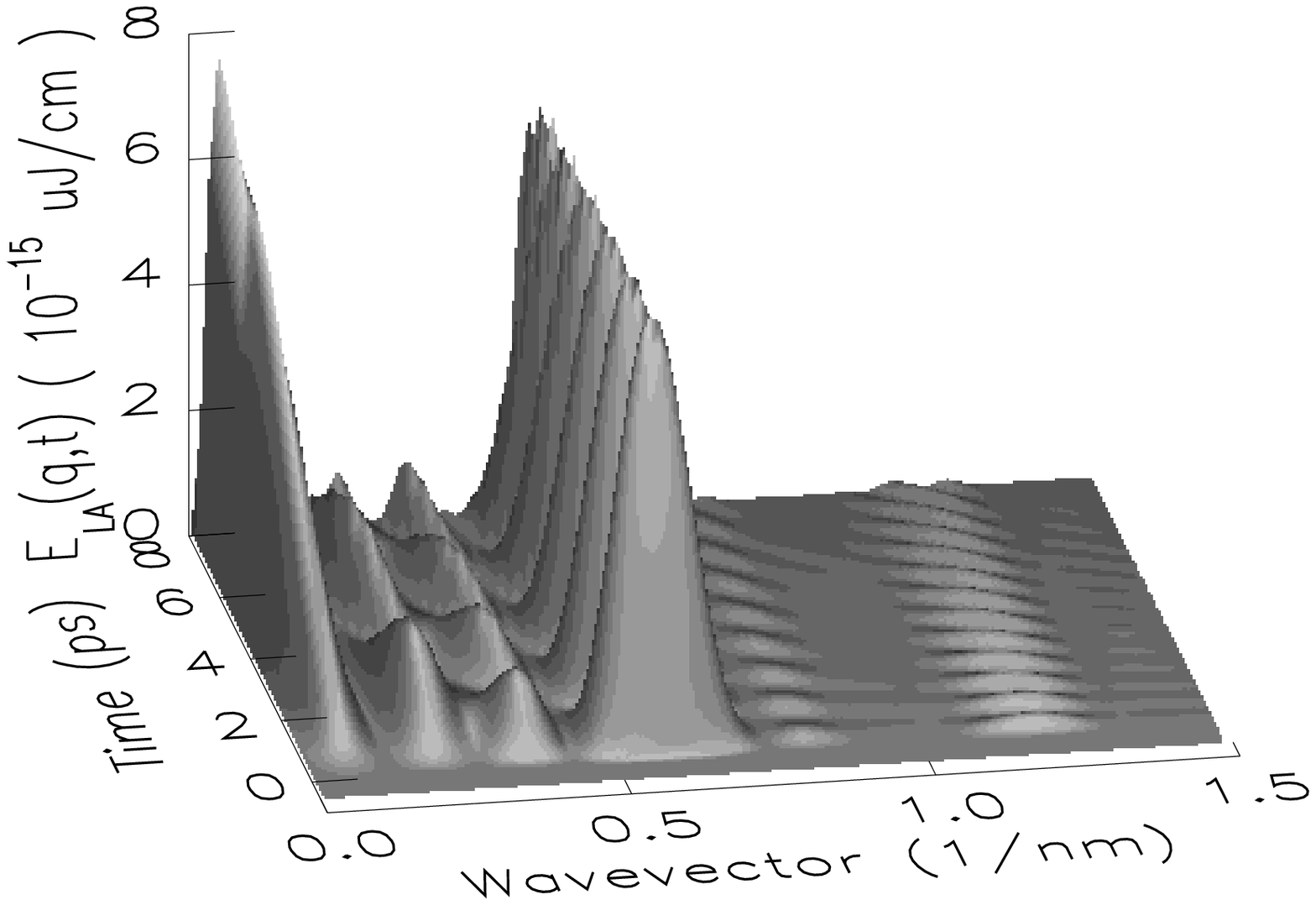}{}
\epsfxsize=3.4in
\epsfbox{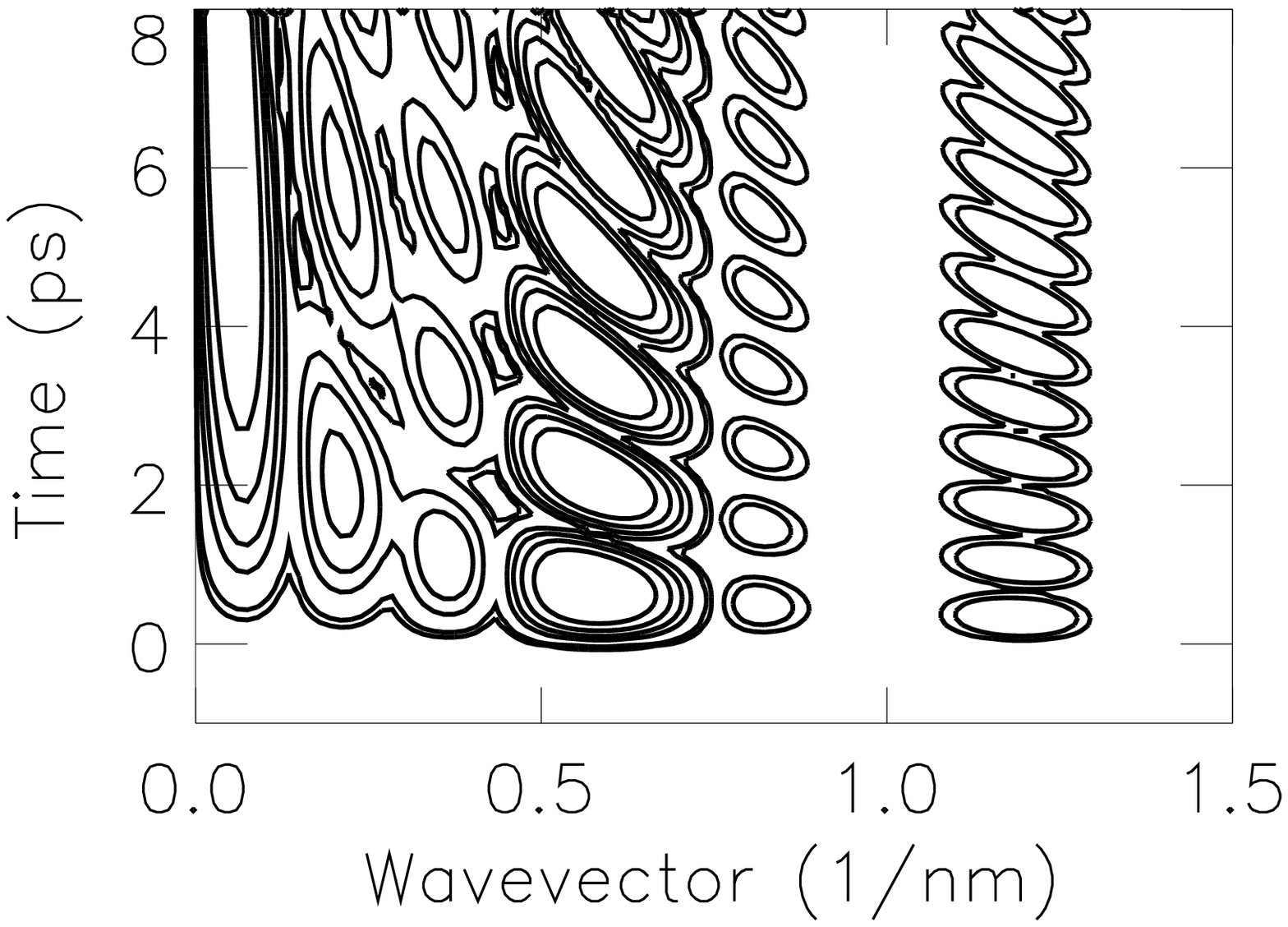}{}
\end{center}
\caption{Total energy density power spectrum, ${\cal{E}}_{LA}(q,t)$,
for coherent LA phonons as
a function of phonon wavevector, $q$, and time for the driving function
Fig. \ref{DrivingFunction}. The total integrated energy density as
a function of time is obtained by integrating over the phonon 
wavevector, $q$.}
\label{EnergyPower}
\end{figure}

The power spectrum of the coherent LA phonon energy density in
$q$-space can be written in terms of the coherent phonon amplitudes
$D_q(t)$. The power spectrum for the total coherent LA phonon energy density,
\begin{mathletters}
\label{DensityEquationsQ}
\begin{equation}
{\cal{E}}_{LA}(q,t)={\cal{T}}_{LA}(q,t)+{\cal{V}}_{LA}(q,t) ,
\label{TotalEnergyDensityQ}
\end{equation}
is again the sum of a kinetic energy term,
\begin{equation}
{\cal{T}}_{LA}(q,t)=\frac{1}{2 A} \ \frac{\hbar}{\omega_q} \
\left| \frac{\partial D_q(t)}{\partial t} \right|^2 ,
\end{equation}
and a potential energy term,
\begin{equation}
{\cal{V}}_{LA}(q,t)=\frac{1}{2 A} \ 
\hbar \omega_q \left|  D_q(t) \right|^2 .
\end{equation}
\end{mathletters}
The phonon energy density per unit area is obtained by summing the
power spectrum over positive phonon wavevectors, $q$. Thus,
\begin{equation}
E_{LA}(t) = \sum_{q>0} \ {\cal{E}}_{LA}(q,t) .
\label{ELAq}
\end{equation}
The total energy density power spectrum for coherent LA phonons as
a function of phonon wavevector, $q$, and time is shown in
Fig. \ref{EnergyPower}.
The peak near $q = 0$ is associated with buildup of the steady state
energy density localized in the MQW region. Secondary peaks are seen
near $q_0=0.59 \ nm^{-1}$ and twice this wavevector, i.e.
$q_1=2 \ q_0=1.18 \ nm^{-1}$. The wavevector $q_0$ corresponds to the
the wavevector of the MQW period,
\begin{equation}
q_0= \frac{2 \pi}{L_w+L_b} ,
\end{equation}  
where $L_w$ and $L_b$ are the well and barrier widths.

\begin{figure}
\begin{center}
\epsfxsize=3.4in
\epsfbox{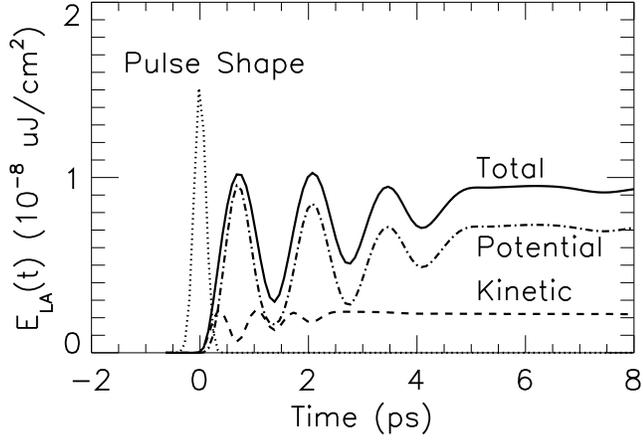}{}
\end{center}
\caption{Integrated energy density, $E_{LA}(t)$, as a function of time
for coherent
LA phonons generated by the driving function shown in
Fig. \ref{DrivingFunction}. The total integrated energy density is the sum
of kinetic- and potential- energy terms. The pulse shape (arb. units) is
shown for comparison. }
\label{TotalEnergy}
\end{figure}

The total coherent LA phonon energy per unit area can be obtained from either
Eqs. (\ref{ELAz}) or (\ref{ELAq}).
In Fig. \ref{TotalEnergy}, we show the total phonon energy per unit
area, $E_{\text{LA}}(t)$, as a function of time for the coherent
LA phonons generated by the driving function shown in
Fig. \ref{DrivingFunction}. The total energy per unit area is the sum of
kinetic- and potential- energy terms. For comparison, the pulse shape is
shown as a dotted line. It is clear from the figure that the buildup of
energy in coherent LA phonons takes place on a time scale that is much
longer than the pump duration. In addition, we see that the total energy
buildup in the phonons saturates at around $5\ \text{ps}$ and that some
strong but rapidly decreasing oscillations are superimposed on top of an
increasing trend.

The saturation phenomenon results from the fact that we have a finite
number of quantum wells and not an infinite superlattice.
The results can best be explained in terms of the driven string equation.
In general, the rate at which energy is fed into the phonon field per
unit area is described by the energy equation, \cite{berg66}
\begin{equation}
\frac{\partial {E}_{\text{LA}}}{\partial t}= \rho_0 \ 
\int_0^L dz\ \ S(z,t) \ V(z,t) ,
\label{EnergyEquation}
\end{equation}
in which $S(z,t)$ and $V(z,t)$ are the phonon driving function and velocity
fields defined in Eqs.\ (\ref{Szt}) and (\ref{Vzt}),
and $\rho_0$ is the $GaN$ mass density used 
for computing the sound speed, $C_s$, in Eq.\ (\ref{PhononDispersion}).
The energy equation simply says that the rate at which energy is added to
a driven string is proportional to the local force times velocity integrated
over the length of the string. The integral in Eq.\ (\ref{EnergyEquation})
vanishes when the transient velocity field, $V(z,t)$, exits the MQW region
in which the driving function, $S(z,t)$, is localized. Thus, the time,
$t_{\text{sat}}$, required for ${\cal{E}}_{LA}$ to saturate is just the
time it takes for  an LA sound wave to cross the MQW, i.e.
$t_{\text{sat}} \approx W/C_s$, where $W$ is the width of the MQW region
over which the driving function is localized. In our example the LA sound 
speed is $C_s = 80 \ \AA / \text{ps}$ in $GaN$ and the MQW width
(four well and three barrier layers) is $W = 381 \AA$, from which we
obtain $t_{\text{sat}}=4.8 \ \text{ps}$.

The oscillations of $E_{LA}$ observed in Fig.\ \ref{TotalEnergy}
reflect the number and periodicity of the diode MQW's. The pump
laser generates spatially periodic electron and hole distributions,
as seen in Fig. \ref{3dDensity1}, due to the fact that the pump 
photoexcites carriers in the wells but not the barriers. From each of 
the wells in the MQW, two sound pulses emerge traveling in
opposite directions thus giving rise to an outwardly propagating velocity field
pattern, $V(z,t)$, with four peaks traveling outward in each direction
as seen in Fig. \ref{UVFigure}. The driving function, $S(z,t)$, on the
other hand, is localized in the MQW's and is relatively constant in time
after the pump pulse dies out. The driving functions localized in each
well do work on four traveling velocity disturbances, the one generated
in the well itself as well as the ones generated in the three neighboring
wells that subsequently pass by. This gives rise to the three peaks plus
saturation plateau seen in Fig. \ref{TotalEnergy}. The time interval between
coresponding peaks in adjacent wells is just the time it takes LA sound waves
to travel between wells, i.e. $t_{\text{period}}=(L_w+L_b)/C_s$, where $L_w$
and $L_b$ are the
well and barrier thicknesses. For the MQW structure, $L_w=63.0 \ \AA$
and $L_b=43.0 \ \AA$ and we have $t_{\text{period}}=1.325 \ \text{ps}$
which agrees with the peak-to-peak time between the first and second
peaks seen in Fig. \ref{TotalEnergy}.

\section{Summary and Conclusions}

We have developed a microscopic theory for the generation and propagation of
coherent LA phonons in pseudomorphically strained wurzite (0001)
$InGaN/GaN$ multi-quantum well (MQW) $pin$ diodes. Both $GaN$ and $InN$ have
different lattice constants so that a significant mismatch occurs between the
wells and barriers. The presence of strain in the MQW's results in the
creation of strain-induced built-in piezoelectric fields on the order of
sevaral MV/cm which significantly alter the electronic and optical properties
of the diode structure. In particular, the effective band gap can be lower
than the band gap in unstrained $InGaN$ wells due to the presence of
triangular piezoelectric potentials.

To a first approximation, the generation of coherent LA phonons is
driven by optical photoexcitation of electron-hole pairs by an
ultrafast Gaussian pump laser. Under typical experimental conditions,
the propagation of coherent LA phonons is described by a \textit{uniform}
loaded string equation for the lattice displacement where the time- and
position-dependent driving force on the string is a function of the
photoexcited carrier density.  This differs from coherent LO phonon
oscillations in bulk systems where the coherent LO phonons obey a forced
oscillator equation.  Both deformation potential and piezoelectric
coupling mechanisms contribute to the driving force. We find that
deformation potential coupling contributes a driving force proportional
to the derivative of the carrier density while piezoelectric coupling
contributes a driving force proportional to the photoexcited carrier
density.

We found that the driving term in the loaded string equation is
suddenly turned on by rapid generation of electron-hole pairs by the
pump and remains approximately constant theafter. This sudden
displaceive loading of the string results in a new static equilibrium
lattice displacement. This new static equilibrium displacement
corresponds to a population of coherent LA phonons with $q \approx 0$.
As the lattice adjusts to the new equilibrium, coherent LA phonons are
transmitted in the positive and negative $z$-directions at the LA sound
speed. These traveling coherent LA phonons are characterized by $q
\approx 2 \pi/ L$ where $L$ is the superlattice period.

The formalism described here can be applied to the analysis of more complicated
device geometries as well as more complicated laser pulse sequences.
This gives a simpler method for calculating the coherent LA phonon generation
in more complicated geometries and gives additional insight
into the acoustic coherent response.

\acknowledgments
The work of GDS and CJS was supported by the National Science Foundation
through Grant No. DMR-9817828 and INT-9414072.
The work of CSK was supported by Chonnam
National University through a grant in the year 1999. 



\begin{table}
\caption{Material parameters for wurtzite InN and GaN. Material parameters
for $In_{x}Ga_{1-x}N$ are obtained through interpolation in $x$ as described
in the text.}
\begin{tabular}{lcr}
Parameter & InN  & GaN \\
\tableline

\underline{Lattice constants}  & & \\
$a_0$( \AA )          & 3.540  \tablenotemark[1]&  3.189 \tablenotemark[1]\\ 
$c_0$( \AA )          & 3.708  \tablenotemark[1]&  5.185 \tablenotemark[1]\\
$u_0$                 & 0.377  \tablenotemark[1]&  0.376 \tablenotemark[1]\\

\\ \underline{Direct band gaps (eV)}  & & \\
$E_g$                 & 1.95   \tablenotemark[2]&  3.40  \tablenotemark[2]\\

\\ \underline{Electron effective masses ($m_0$)} & & \\
$m^{*}_{xy}$          & 0.10   \tablenotemark[3]&  0.18 \tablenotemark[3]\\
$m^{*}_{z}$           & 0.11   \tablenotemark[3]&  0.19 \tablenotemark[3]\\  

\\ \underline{Hole effective mass parameters}  & & \\
$A_1$                 &  -9.28 \tablenotemark[3]& -7.24 \tablenotemark[3]\\
$A_2$                 &  -0.60 \tablenotemark[3]& -0.51 \tablenotemark[3]\\
$A_3$                 & \ 8.68 \tablenotemark[3]&  6.73 \tablenotemark[3]\\
$A_4$                 &  -4.34 \tablenotemark[3]& -3.36 \tablenotemark[3]\\
$A_5$                 &  -4.32 \tablenotemark[3]& -3.35 \tablenotemark[3]\\
$A_6$                 &  -6.08 \tablenotemark[3]& -4.72 \tablenotemark[3]\\

\\ \underline{Hole splitting energies (meV)}  & & \\
$\Delta_1=\Delta_{cr}$   & 17.0 \tablenotemark[3]& 22.0 \tablenotemark[3]\\
$\Delta_2=\Delta_{s0}/3$ & 1.0  \tablenotemark[3]& 3.67 \tablenotemark[3]\\
$\Delta_3$               & 1.0  \tablenotemark[3]& 3.67 \tablenotemark[3]\\

\\ \underline{Electron deformation potentials (eV)}  & & \\
$a_{c,xy}$            &                          & -4.08 \tablenotemark[4]\\
$a_{c,z} $            &                          & -4.08 \tablenotemark[4]\\

\\ \underline{Hole deformation potentials (eV)}  & & \\
$D_1$                 &                          &  0.7  \tablenotemark[4]\\
$D_2$                 &                          &  2.1  \tablenotemark[4]\\
$D_3$                 &                          &  1.4  \tablenotemark[4]\\
$D_4$                 &                          & -0.7  \tablenotemark[4]\\

\\ \underline{Piezoelectric constants (C/$m^2)$}  & & \\
$e_{31}$              &  -0.57  \tablenotemark[1]&  -0.49 \tablenotemark[1]\\ 
$e_{33}$              & \ 0.97  \tablenotemark[1]&   0.73 \tablenotemark[1]\\

\\ \underline{Elastic stiffness constants (GPa)}  & & \\
$C_{11}$              & 190  \tablenotemark[1]&  374 \tablenotemark[1]\\ 
$C_{12}$              & 104  \tablenotemark[1]&  106 \tablenotemark[1]\\
$C_{13}$              & 121  \tablenotemark[1]&   70 \tablenotemark[1]\\
$C_{33}$              & 182  \tablenotemark[1]&  379 \tablenotemark[1]\\
$C_{44}$              & \ 10 \tablenotemark[1]&  101 \tablenotemark[1]\\

\\ \underline{Static dielectric constant}  & & \\
$\varepsilon_0$       & 15.3  \tablenotemark[5]&  8.9 \tablenotemark[6] 

\end{tabular}
\label{ParameterTable}

\tablenotetext[1] {Ref. \ \onlinecite{ambacher99}. }
\tablenotetext[2] {Ref. \ \onlinecite{nakamura97}. }
\tablenotetext[3] {Ref. \ \onlinecite{yeo98}.      }
\tablenotetext[4] {Ref. \ \onlinecite{chuang96b}.  }
\tablenotetext[5] {Ref. \ \onlinecite{martin96}. }
\tablenotetext[6] {Ref. \ \onlinecite{doshi98}. }
\end{table}

\begin{table}
\caption{Simulation parameters for photogeneration of coherent acoustic
phonons in a four well MQW diode under flat band biasing conditions.
A schematic of the diode structure is shown in Fig.\ \ref{diode} . }
\begin{tabular}{lc}

\underline{MQW diode structure}  &  \\

Left $GaN$ buffer width (\AA)          &      43.0    \\
\\
Number of wells                        &      4       \\
Well width (\AA)                       &      63.0    \\
Indium fraction in well                &      0.06    \\
$GaN$ barrier width (\AA)              &      43.0    \\
\\
Right $GaN$ buffer width (\AA)         &      43.0    \\

\\ \underline{Applied bias}  &  \\
$V_A$ ($V$)                            &    -0.261    \\

\\ \underline{Lattice temperature}  &  \\
$T$ ($K$)                              &     300.0    \\

\\ \underline{Pump parameters}  &  \\
Photon energy (eV)                     &      3.21    \\
Fluence ( $\mu J$/$cm^2$ )             &     160.0    \\
Gaussian FWHM ($fs$)                   &     180.0    \\
Polarization                           & Left circular         
\end{tabular}
\label{SimulationTable}
\end{table}


\begin{references}
\bibitem{note1}
This will affect the forcing term, $S(z,t)$, in Eq. ( \ref{Szt} ).
\end{references}
\end{document}